\documentclass[sigconf]{acmart}
\usepackage{bm}
\usepackage{url}
\DeclareUnicodeCharacter{2212}{-}
\usepackage{subcaption}
\copyrightyear{2021}
\acmYear{2021}
\setcopyright{acmcopyright}
\acmConference[MM '21] {Proceedings of the 29th ACM International Conference on Multimedia}{October 20--24, 2021}{Virtual Event, China.}
\acmBooktitle{Proceedings of the 29th ACM Int'l Conference on Multimedia (MM '21), Oct. 20--24, 2021, Virtual Event, China}
\acmPrice{15.00}
\acmDOI{10.1145/3474085.3475213}
\acmISBN{978-1-4503-8651-7/21/10}

\begin{document}
\fancyhead{}

\title{Enhanced Invertible Encoding for Learned Image Compression}

\author{Yueqi Xie}
\authornote{Both authors contributed equally.}
\affiliation{%
  \institution{HKUST}
  \country{ }
}

\author{Ka Leong Cheng}
\authornotemark[1]
\affiliation{%
  \institution{HKUST}
  \country{ }
}

\author{Qifeng Chen}
\affiliation{%
  \institution{HKUST}
  \country{ }
}




\begin{abstract}
Although deep learning based image compression methods have achieved promising progress these days, the performance of these methods still cannot match the latest compression standard Versatile Video Coding (VVC). Most of the recent developments focus on designing a more accurate and flexible entropy model that can better parameterize the distributions of the latent features. However, few efforts are devoted to structuring a better transformation between the image space and the latent feature space. In this paper, instead of employing previous autoencoder style networks to build this transformation, we propose an enhanced Invertible Encoding Network with invertible neural networks (INNs) to largely mitigate the information loss problem for better compression. Experimental results on the Kodak, CLIC, and Tecnick datasets show that our method outperforms the existing learned image compression methods and compression standards, including VVC (VTM 12.1), especially for high-resolution images. Our source code is available at~\url{https://github.com/xyq7/InvCompress}.
\end{abstract}




\maketitle

\section{Introduction}
Lossy image compression has been a fundamental and important research topic in media storage and transmission for decades. Classical image compression standards are usually based on handcrafted schemes, such as JPEG~\cite{wallace1992jpeg}, JPEG2000~\cite{rabbani2002jpeg2000}, WebP~\cite{webp}, BPG~\cite{bpg}, and Versatile Video Coding (VVC)~\cite{vvc}. Some of them are widely used in practice. Recently, there is an increasing interest in learned image compression methods~\cite{balle2016end, minnen2018joint, mentzer2018conditional, liu2019non, cheng2020learned} considering their competitive performance.
Generally, the recent VAE-based methods~\cite{balle2016end, Ball2017EndtoendOI} follow a process: an encoder transforms the original pixels $\mathbf{x}$ into lower-dimensional latent features $\mathbf{y}$ and then quantizes it to $\mathbf{\hat{y}}$ , which can be losslessly compressed using entropy coding methods, like arithmetic coding~\cite{arith, Rissanen1981Universal}. A jointly optimized decoder is utilized to transform $\mathbf{\hat{y}}$ back to the reconstructed original image $\mathbf{\hat{x}}$. 

Most of the recent developments on this topic focus on the improvement of the entropy model. Ball\'{e} et al.~\cite{balle2018variational} propose a variational image compression method with a scale hyperprior. Afterward, Minnen et al.~\cite{minnen2018joint} combine the context model~\cite{lee2018context} and an autoregressive model over the latent features. Cheng et al.~\cite{cheng2020learned} further improve the method with attention modules and employ discretized Gaussian mixture likelihoods to better parameterize the latent features. These recent improvements on entropy models greatly contribute to efficient compression. However, they usually model the transformation between the image space and the latent feature space using an autoencoder framework. Although autoencoders have strong capacities to select the important portions of the information for reconstruction, the neglected information during encoding is usually lost and unrecoverable for decoding. 

To resolve the information loss problem, a possibly good choice is to integrate the idea of invertible neural networks (INNs) into the encoding-decoding process because INNs have the strictly invertible property to help preserve information. However, it is still nontrivial to leverage INNs for replacing the original encoder-decoder architecture. On the one hand, to ensure strict invertibility, the total pixel number in the output should be the same as the input pixel number. 
It is intractable to quantize and compress such high-dimensional features~\cite{gersho2012vector} and hard to get desirable low bit rates for compression. The work~\cite{helminger2021lossy} explores the usage of INN to break AE limit but only gets good performance in high bpp range.
In the work~\cite{wang2020modeling} attempting to use INNs for lower-bpp image compression, they take a subspace of the output to get lower-dimensional features, model lost information with a given distribution, and do sampling during the inverse passing. However, this strategy introduces unwanted errors and leads to unstable training, so they further introduce a knowledge distillation module to guide and stabilize the training of INNs, but it may lead to sub-optimal solutions. 
On the other hand, although the mathematical network design guarantees the invertibility of INNs, it makes INNs have limited nonlinear transformation capacity~\cite{dinh2015nice} compared to other encoder-decoder architectures.

Concerning these aspects, we propose an enhanced Invertible Encoding Network for image compression, which maintains a highly invertible structure based on INN. Instead of using the unstable sampling mechanism for training~\cite{wang2020modeling}, we propose an attentive channel squeeze layer to stabilize the training and to flexibly adjust the feature dimension for a lower bit rate. We also present a feature enhancement module to improve the network nonlinear representation capacity, where the same-resolution transformation and the residual connection of this module help preserve image information. With our design, we can directly integrate INNs to capture the lost information without relying on any sub-optimal guidance and boost performance.

Experimental results show that our proposed method outperforms the existing learned methods and traditional codecs on three widely-used datasets for image compression, including Kodak~\cite{kodak} and two high-resolution datasets, namely the CLIC Professional Validation dataset~\cite{CLIC2020} and the Tecnick dataset~\cite{asuni2014tecnick}. Visual results demonstrate that our reconstructed images contain more details under similar BPP rates, beneficial from our highly invertible design. Further analysis of our proposed attentive channel squeeze layer and feature enhancement module verifies their functionality. The contributions of this paper can be summarized as follows:
\begin{itemize}
    \item Unlike widely-used autoencoder style networks for learned image compression, we present an enhanced Invertible Encoding Network with an INN architecture for the transformation between image space and feature space to largely mitigate the information loss problem.
    \item The proposed method outperforms the existing learned image compression methods and traditional compression codecs, including the latest VVC (VTM 12.1), especially on two high-resolution datasets. 
    \item We propose an attentive channel squeeze layer to resolve the unstable and sub-optimal training of the INN-based networks for image compression. We further leverage a feature enhancement module to improve the nonlinear representation capacity of our network.
\end{itemize}

\section{Related Work}
\subsection{Lossy Image Compression}
\textbf{Traditional methods.} Lossy image compression has long been an important and fundamental topic in image processing. Traditional compression standards includes JPEG~\cite{wallace1992jpeg}, JPEG2000~\cite{rabbani2002jpeg2000}, WebP~\cite{webp}, Better Portable Graphics (BPG)~\cite{bpg} and Versatile Video Coding (VVC)~\cite{vvc}. Many of them are widely used in practice. Typically, they follow the pipeline of transformation, quantization, and entropy coding. The transformation is usually based on handcrafted modules with prior knowledge, such as discrete cosine transformation (DCT)~\cite{ahmed1974dct} and discrete wavelet transform (DWT)~\cite{marpe2003context}. The entropy coders include Huffman coder and some arithmetic coding methods~\cite{Rissanen1981Universal, arith, marpe2003context}. Some modern standards like BPG and VVC further introduce intra prediction for better compression performance. However, since these traditional methods generally perform compression based on image blocks, their reconstructed images usually contain some limitations of the blocking effects.

\textbf{Learned methods.} In these years, deep learning based methods have raised great interest with impressive performance. These methods try to employ neural networks instead of handcrafted rules for the nonlinear transformation learning between the image space and the latent feature space. 

Several recurrent neural network (RNN) based methods~\cite{toderici2015variable, toderici2017full, johnston2018improved} progressively encode the residual information from the previous step to compress the image. However, these RNN-based models rely on binary representation at each iteration and cannot directly optimize the rate during training. 

Another branch of the methods is based on variational autoencoders (VAEs). Several early works~\cite{Theis2017LossyIC, Ball2017EndtoendOI, Agustsson2017SofttoHardVQ} solve the problem of non-differential quantization and estimation of bit rates, which lay the foundation for end-to-end optimization to minimize the estimated bit rates and the reconstructed image distortion. Afterward, the improvement is mainly in two directions. One direction is to build a more effective entropy model for rate estimation. The work~\cite{balle2018variational} proposes a hyperprior entropy model to use additional bits to capture the distribution of the latent features. Some follow-up methods~\cite{minnen2018joint, lee2018context, mentzer2018conditional} integrate context factors to further reduce the spatial redundancy within the latent features. Also, 3D context entropy model~\cite{guo2020context}, channel-wise models~\cite{MinnenS20} and hierarchical entropy models~\cite{minnen2018joint, hu2020coarse} are 
used to better extract correlations of latent features. Cheng et al.~\cite{cheng2020learned} propose Gaussian mixture likelihood to replace the original single Gaussian model to improve the accuracy. Another direction is to improve the VAE architecture. CNNs with generalized divisive normalization (GDN) layers~\cite{balle2016end, Ball2017EndtoendOI} achieve good performance for image compression. Attention mechanism and residual blocks~\cite{liu2019non, zhou2019end, zhang2019residual, cheng2020learned} are incorporated into the VAE architecture. Some other progress includes generative adversarial training ~\cite{rippel2017real,santurkar2017generative, agustsson2019generative}, spatial RNNs~\cite{lin2020} and multi-scale fusion~\cite{rippel2017real}.

Generally, the VAE-based methods account for a more significant proportion among all the existing learned methods for image compression, considering the great performance and stability of the encoder and decoder architecture. Still, they cannot explicitly solve the information loss problem during encoding, making the neglected information usually unrecoverable for decoding.

\subsection{Invertible Neural Networks}
Invertible neural networks (INNs)~\cite{dinh2015nice, dinh2017density, kingma2018glow} are popular with three great properties of the design: (i) INNs maintain a bijective mapping between inputs and outputs; (ii) the bijective mapping is efficiently accessible; (iii) there exists a tractable Jacobian of the bijective mapping to compute posterior probabilities explicitly. Many recent works in different areas with INN architecture achieve better performance than those with autoencoder style frameworks, especially for the tasks with inherent invertibility.

RealNVP~\cite{dinh2017density} uses a multi-scale architecture with coupling layers and convolution operations to first deal with image processing tasks. Ardizzone et al.~\cite{ardizzone2018analyzing} demonstrate the effectiveness of INNs on both the synthetic data and two real-world applications in the medicine and astrophysics fields. Recently, many works start to use normalizing flow methods with exact likelihoods during training in generative tasks, where the key is to parametrize the distribution using INNs. SRFlow~\cite{lugmayr2020srflow} uses a conditional INN architecture to better resolve the ill-posed problem of super-resolution compared to GAN-based methods. Pumarola et al.~\cite{pumarola2020cflow} propose a conditional generative flow model for Image and 3D Point Cloud generation. For the image rescaling task, Xiao et al.~\cite{xiao2020invertible} use INNs to generate a bijective mapping between high-resolution images and low-resolution images with additional latent variables. Xing et al.~\cite{Xing2021} propose an invertible image signal processing pipeline based on INNs.

\begin{figure*}[t]
\begin{subfigure}{0.14\linewidth}
    \centering
    \includegraphics[width=1.0\linewidth]{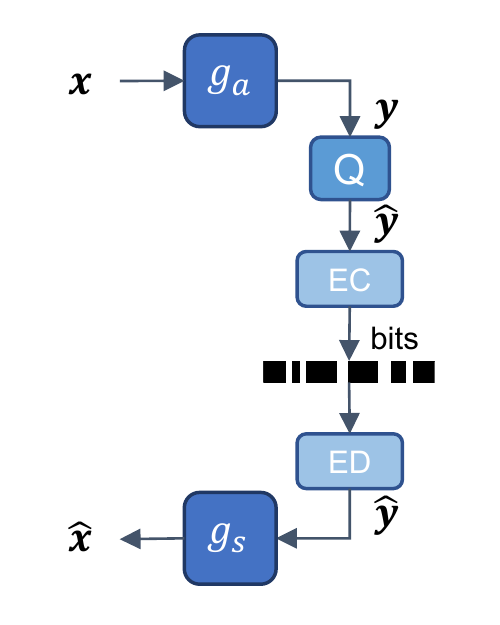}
    \subcaption{Baseline~\cite{balle2016end}}
    \label{fig:history-a}
\end{subfigure}
\begin{subfigure}{0.28\linewidth}
    \centering
    \includegraphics[width=1.0\linewidth]{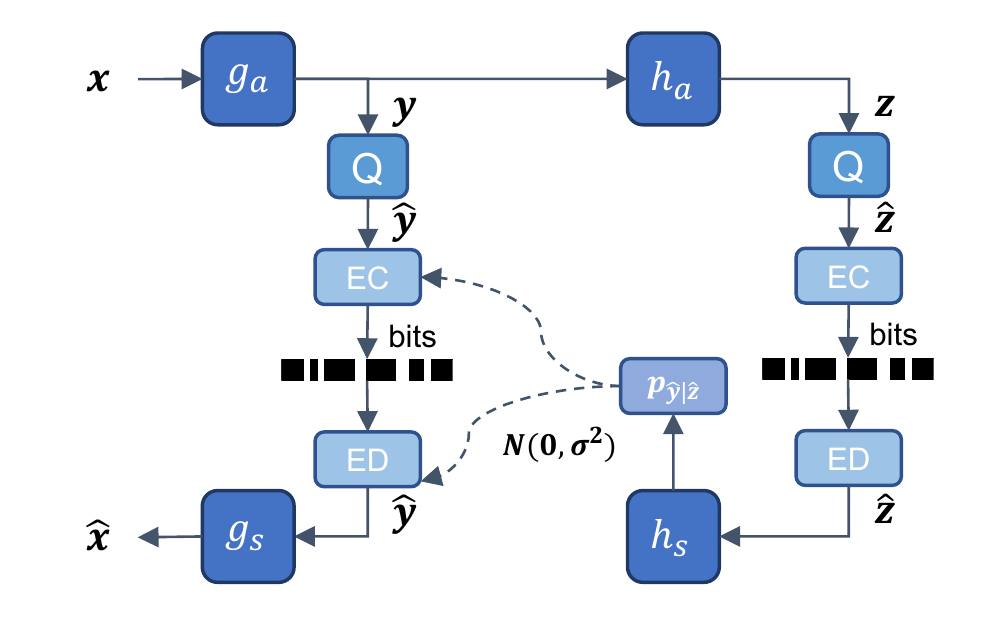}
    \subcaption{Ball\'{e} et al.~\cite{balle2018variational}}
    \label{fig:history-b}
\end{subfigure}
\begin{subfigure}{0.28\linewidth}
    \centering
    \includegraphics[width=1.0\linewidth]{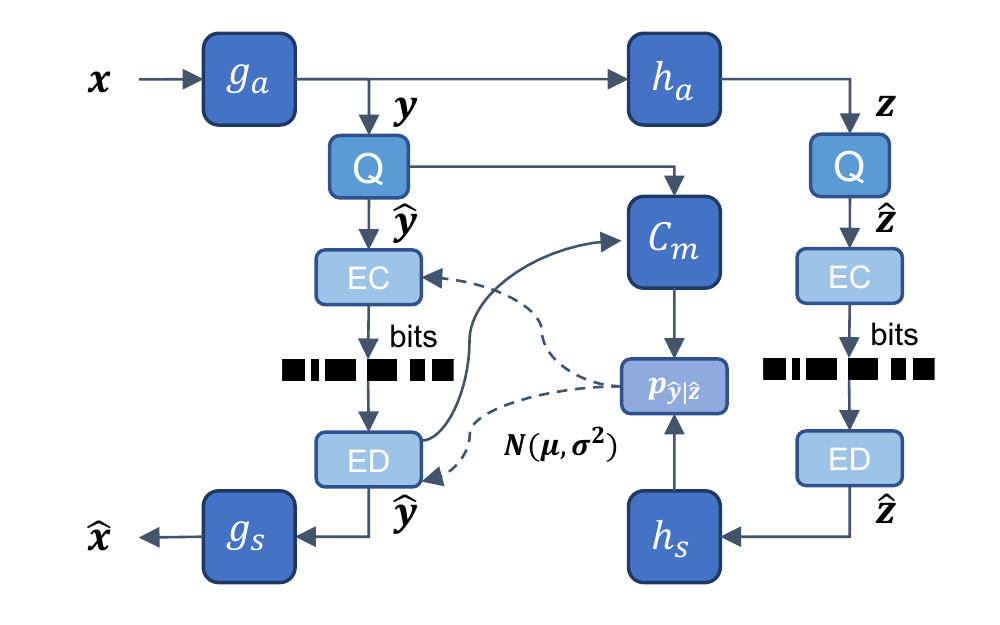}
    \subcaption{Minnen et al.~\cite{minnen2018joint}}
    \label{fig:history-c}
\end{subfigure}
\begin{subfigure}{0.28\linewidth}
    \centering
    \includegraphics[width=1.0\linewidth]{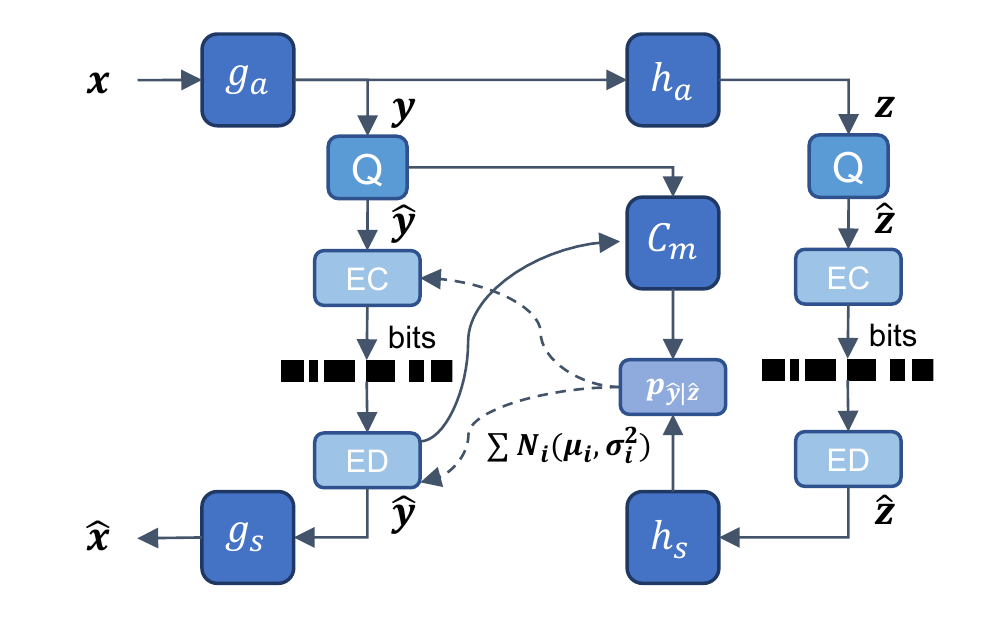}
    \subcaption{Cheng et al.~\cite{cheng2020learned}}
    \label{fig:history-d}
\end{subfigure}
\caption{The overview of existing learned compression methods. $EC$ and $ED$ denote the encoder and decoder of the entropy model, respectively, and $C_m$ denotes the context model.}
\label{fig:hitory}
\end{figure*}

\section{Method}
\subsection{Background} 
Figure~\ref{fig:hitory} provides a high-level overview of general learned image compression models in the transform coding approach~\cite{Goyal2001Theoretical}. A baseline model (Figure~\ref{fig:history-a}) is formulated as follows:
\begin{equation}
    \mathbf{y} = g_a(\mathbf{x}),\ \mathbf{\hat{y}} = Q(\mathbf{y}),\ \mathbf{\hat{x}} = g_s(\mathbf{\hat{y}}).
\end{equation}
For encoding, a parametric analysis transform $g_a$ encodes the image $\mathbf{x}$ into latent features $\mathbf{y}$, and $\mathbf{y}$ is then quantized to obtain the discrete latent features $\mathbf{\hat{y}}$, which is then losslessly compressed into bitstreams using entropy coding like arithmetic coding~\cite{Rissanen1981Universal}. Note that since integer rounding is a fundamentally non-differentiable function, we use the method in~\cite{Ball2017EndtoendOI} to approximately model the quantized latent features by adding a uniform noise $U(-0.5, 0.5)$ to $\mathbf{y}$ during training. For simplicity, we use $\mathbf{\hat{y}}$ to denote both the latent features with uniform noise added during training and the discretely quantized latent features during testing. For decoding, a parametric synthesis transform $g_s$ decodes the quantized $\mathbf{\hat{y}}$ back to the reconstructed image $\mathbf{\hat{x}}$.

The fundamental optimization objective of learned image compression is to minimize a weighted sum of the rate-distortion tradeoff during training: 
\begin{equation}
    L = R(\mathbf{\hat{y}}) + \lambda D(\mathbf{x}, \mathbf{\hat{x}}).
\end{equation}
The rate $R$ is the the entropy of $\mathbf{\hat{y}}$, which is estimated by a non-parametric, fully factorized density model (entropy model) $p_{\mathbf{\hat{y}}|\theta}$ during training. So, we have the rate estimation:
\begin{equation}
    R = \mathbb{E}[-\log_2 p_{\mathbf{\hat{y}}|\theta}(\mathbf{\hat{y}}|\theta)].
\end{equation}
The distortion $D$ is defined as $D = MSE(\mathbf{x}, \mathbf{\hat{x}})$ for MSE optimization and $D = 1 − MS\text{-}SSIM(\mathbf{x}, \mathbf{\hat{x}})$ for MS-SSIM~\cite{wang2003msssim} optimization. We use $\lambda$ to control the rate-distortion tradeoff for different bit rates.

Later, Ball\'{e} et al.~\cite{balle2018variational} propose a scale hyperprior on top of the general learn image compression model, as shown in Figure~\ref{fig:history-b}. Specifically, they stack another parametric analysis transform $h_a$ on top of $\mathbf{y}$ to capture the significant spatial dependencies in the quantized latent features $\mathbf{\hat{y}}$ and obtain an additional set of encoded variables $\mathbf{z}$. In this way, they model $\mathbf{\hat{y}}$ as a zero-mean Gaussian distribution with standard deviations $\mathbf{\sigma}$ for all the elements. The standard deviations are estimated by another parametric synthesis transform $h_s$, which takes the quantized $\mathbf{\hat{z}}$ as input and outputs the estimated standard deviations $\mathbf{\hat{\sigma}}$. So, we have the conditional probability distribution $p_{\mathbf{\hat{y}}|\mathbf{\hat{z}}} = \mathcal{N}(0, \mathbf{\sigma}^2)$. Similarly, an entropy model $p_{\mathbf{\hat{z}}|\theta}$ is applied for entropy estimation of $\mathbf{\hat{z}}$. In this case, the rate estimation contains two terms:
\begin{equation}
    R = \mathbb{E}[-\log_2 p_{\mathbf{\hat{y}}|\mathbf{\hat{z}}}(\mathbf{\hat{y}}|\mathbf{\hat{z}})] + \mathbb{E}[-\log_2 p_{\mathbf{\hat{z}}|\theta}(\mathbf{\hat{z}}|\theta)].
\end{equation}

Later works further improve the hyperprior to better parameterize the distributions of the quantized latent features with a more accurate and flexible entropy model. Minnen~\cite{minnen2018joint} propose an autoregressive context model with a mean and scale hyperprior, as shown in Figure~\ref{fig:history-c}. Cheng et al.~\cite{cheng2020learned} utilize discretized Gaussian mixture likelihoods with attention enhancement to model the distributions of $\mathbf{\hat{y}}$, as shown in Figure~\ref{fig:history-d}.

\begin{figure*}[t]
\begin{center}
\includegraphics[width=0.99\linewidth]{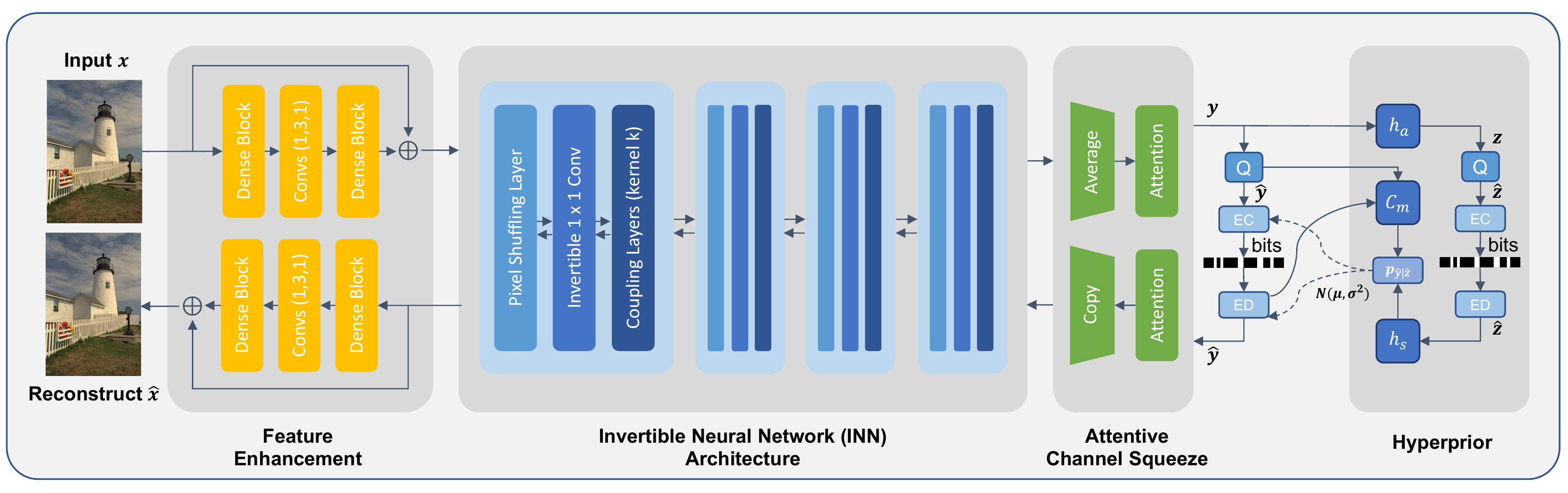}
\end{center}
\caption{Overview and workflow of our proposed enhanced Invertible Encoding Network.}
\label{fig:overview}
\end{figure*}

\subsection{Proposed Method}
Instead of optimizing the parameterization $h_a$, $h_s$ of the latent feature distribution, our proposed method focuses on enhancing the analysis $g_a$ and synthesis $g_s$ transforms between the image space $\mathcal{X}$ and the latent feature space $\mathcal{Y}$. Considering the natural invertibility in image compression, we use an invertible network design with a feature enhancement module and an attentive channel squeeze layer to play the role of the analysis $g_a$ and synthesis $g_s$ transforms. Figure~\ref{fig:overview} shows an overview of our proposed approach.

\textbf{INN architecture.} We formulate an INN architecture design to serve as the analysis $g_a$ and synthesis $g_s$ transforms. It consists of two essential invertible layers: the downscaling layer and the coupling layer. Existing methods~\cite{balle2018variational, cheng2020learned, minnen2018joint} usually employ $4$ downscaling and $4$ corresponding upscaling modules in the analysis and synthesis transforms, respectively. Similarly, we also stack $4$ invertible blocks in our INN architecture to downscale the input resolution by a factor of $2^4$, where each block sequentially contains $1$ downscaling layer and $3$ coupling layers. The kernel sizes of the coupling layers for the $4$ blocks are empirically set as $k=5,5,3,3$.

The downscaling layer is composed of a pixel shuffling layer~\cite{dinh2017density} and an invertible $1 \times 1$ convolution~\cite{kingma2018glow}, and each downscaling layer reduces the resolution of the input tensor by $2$ and quadruples the channel dimension.

We use the affine coupling layer design introduced in RealNVP~\cite{dinh2017density}. For the $i^{th}$ coupling layer that takes an input $\mathbf{u}^{(i)}_{1:C}$ with dimensional size of $C$, it splits the input at position $c < C$ into two parts and gives a $C$ dimensional output $\mathbf{u}^{(i+1)}_{1:C}$:
\begin{equation}
    \mathbf{u}^{(i+1)}_{1:c} = \mathbf{u}^{(i)}_{1:c} \odot \exp(\sigma_c(g_2(\mathbf{u}^{(i)}_{c+1:C}))) + h_2(\mathbf{u}^{(i)}_{c+1:C}),
\end{equation}
\begin{equation}
    \mathbf{u}^{(i+1)}_{c+1:C} = \mathbf{u}^{(i)}_{c+1:C} \odot \exp(\sigma_c(g_1(\mathbf{u}^{(i+1)}_{1:c}))) + h_1(\mathbf{u}^{(i+1)}_{1:c}),
\end{equation}
where $\odot$ denotes the Hadamard product, $\exp(\cdot)$ denotes the exponential function, and $\sigma_c(\cdot)$ denotes the center sigmoid function. Symmetrically, the $i^{th}$ coupling layer inversely takes $\mathbf{u}^{(i+1)}_{1:C}$ as input with splitting position $c$. The affine coupling layer gives a perfect inverse:
\begin{equation}
    \mathbf{u}^{(i)}_{c+1:C} = (\mathbf{u}^{(i+1)}_{c+1:C} - h_1(\mathbf{u}^{(i+1)}_{1:c})) \odot \exp(-\sigma_c(g_1(\mathbf{u}^{(i+1)}_{1:c}))),
\end{equation}
\begin{equation}
    \mathbf{u}^{(i)}_{1:c} = (\mathbf{u}^{(i+1)}_{1:c} - h_2(\mathbf{u}^{(i)}_{c+1:C})) \odot \exp(-\sigma_c(g_2(\mathbf{u}^{(i)}_{c+1:C}))).
\end{equation}
Please note that such invertibility is inherently guaranteed by the mathematical design. Thus, the invertibility holds for any arbitrary feedforward functions $g_1, g_2, h_1, h_2$, meaning that these functions need not be invertible. In our implementations, all these four functions use the following bottleneck design: Conv-LeakyReLU-Conv-LeakyReLU-Conv, where the kernel size of the first and the last convolution is set as $k$; the middle one is a convolutional layer with filter size $1$.

\textbf{Feature enhancement module.} INNs are powerful when modeling invertible transformation. However, since the property of invertibility is guaranteed by the strictly invertible network design, INNs usually have a limited capacity of nonlinear representation~\cite{dinh2015nice}. Hence, we add a feature enhancement module in a residual manner before the INN architecture to improve the nonlinear representativeness of our network. Specifically, this module is based on the popular Dense Block~\cite{huang2017dbnet}, and ``Convs (1,3,1)'' means three cascade convolutions with kernel size $1$, $3$, $1$.

\textbf{Attentive channel squeeze.} All the operations in INNs cannot change the total number of pixels in the input tensor, many of which are simply redundant pixels to be compressed. To resolve this problem, we introduce an attentive channel squeeze layer to reduce the channel dimension of the output tensor of INNs. The idea is as follows: Given a compression ratio $\alpha$ and an input tensor $\mathbf{v}$ with size ($C$, $H$, $W$), the attentive channel squeeze layer first forwardly reshape the tensor into a shape of ($\alpha$, $\frac{C}{\alpha}$, $H$, $W$). Then it performs average operation along the first dimension to obtain the latent features $\mathbf{y}$ with size ($\frac{C}{\alpha}$, $H$, $W$), followed by an attention module proposed in~\cite{cheng2020learned}. For the inverse process, the attentive channel squeeze layer makes $\alpha$ copies of the quantized $\mathbf{\hat{y}}$ first after the attention module and reshapes it into a size of ($C$, $H$, $W$). 

\subsection{Overall Workflow}
For the encoding process, given an input image $\mathbf{x} \in \mathbb{R}^{3 \times H \times W}$ to be compressed with height $H$ and width $W$, the feature enhancement module first extracts and adds some nonlinear representation to obtain $\mathbf{u} \in \mathbb{R}^{3 \times H \times W}$. Then the forward pass of our INN architecture transforms $\mathbf{u}$ into $\mathbf{v} \in \mathbb{R}^{d \times h \times w}$, where $d = 3 \times 4^4$, $h = \frac{H}{2^4}$, and $w = \frac{W}{2^4}$ under our architecture design. Given a hyper-parameter compression ratio $\alpha$, the attentive channel squeeze layer further compresses $\mathbf{v}$ into latent features $\mathbf{y} \in \mathbb{R}^{\frac{d}{\alpha} \times h \times w}$.

For the decoding process, the attentive channel squeeze layer copies the quantized latent features $\mathbf{\hat{y}}$ after attention enhancement for $\alpha$ times and then reshapes it as $\mathbf{\hat{v}}$. The inverse pass of the INN architecture decodes $\mathbf{\hat{v}}$ to obtain $\mathbf{\hat{u}}$, which finally undergoes the feature enhancement module for the reconstructed image $\mathbf{\hat{x}}$.

We employ the same hyperprior proposed by Minnen et al.~\cite{minnen2018joint}, which uses a mean and scale Gaussian distribution to parameterize the quantized latent features $\mathbf{\hat{y}}$ with a pair of analysis $h_a$ and synthesis $h_s$ transforms. Specifically, $h_a$ obtains the side information $\mathbf{z} = h_a(\mathbf{\hat{y}})$, and $h_s$ takes the quantized side information $\mathbf{\hat{z}}$ as input. The output $h_s(\mathbf{\hat{z}})$ works with the causal context $C_m(\mathbf{\hat{y}})$ for the mean and scale Gaussian model $p_{\mathbf{\hat{y}}|\mathbf{\hat{z}}} \leftarrow h_s(\mathbf{\hat{z}}), C_m(\mathbf{\hat{y}})$. Since there is no prior for $\mathbf{\hat{z}}$, a factorized-prior entropy model $\theta$ introduced in~\cite{Ball2017EndtoendOI} is used to parametize the distribution of $\mathbf{\hat{z}}$ as $p_{\mathbf{\hat{z}}|\theta}$. For entropy coding, we use the asymmetric numeral systems (ANS)~\cite{duda2009asymmetric} to losslessly compress $\mathbf{\hat{y}}$ and $\mathbf{\hat{z}}$ into bitstreams.

\begin{figure*}[t]
    \begin{subfigure}{0.5\linewidth}
        \includegraphics[width=.96\linewidth]{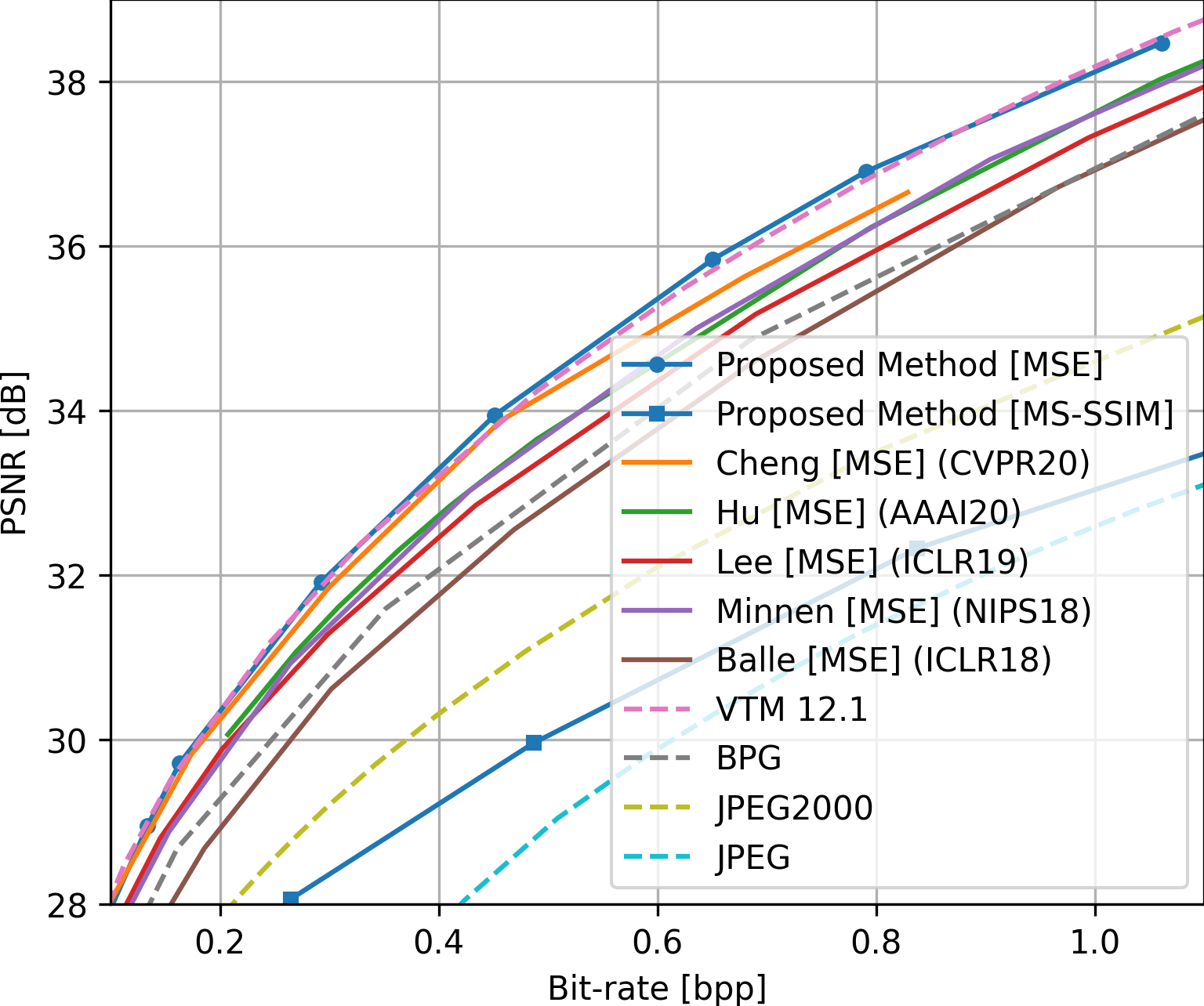}
    \end{subfigure}%
    \begin{subfigure}{0.5\linewidth}
        \centering
        \includegraphics[width=.96\linewidth]{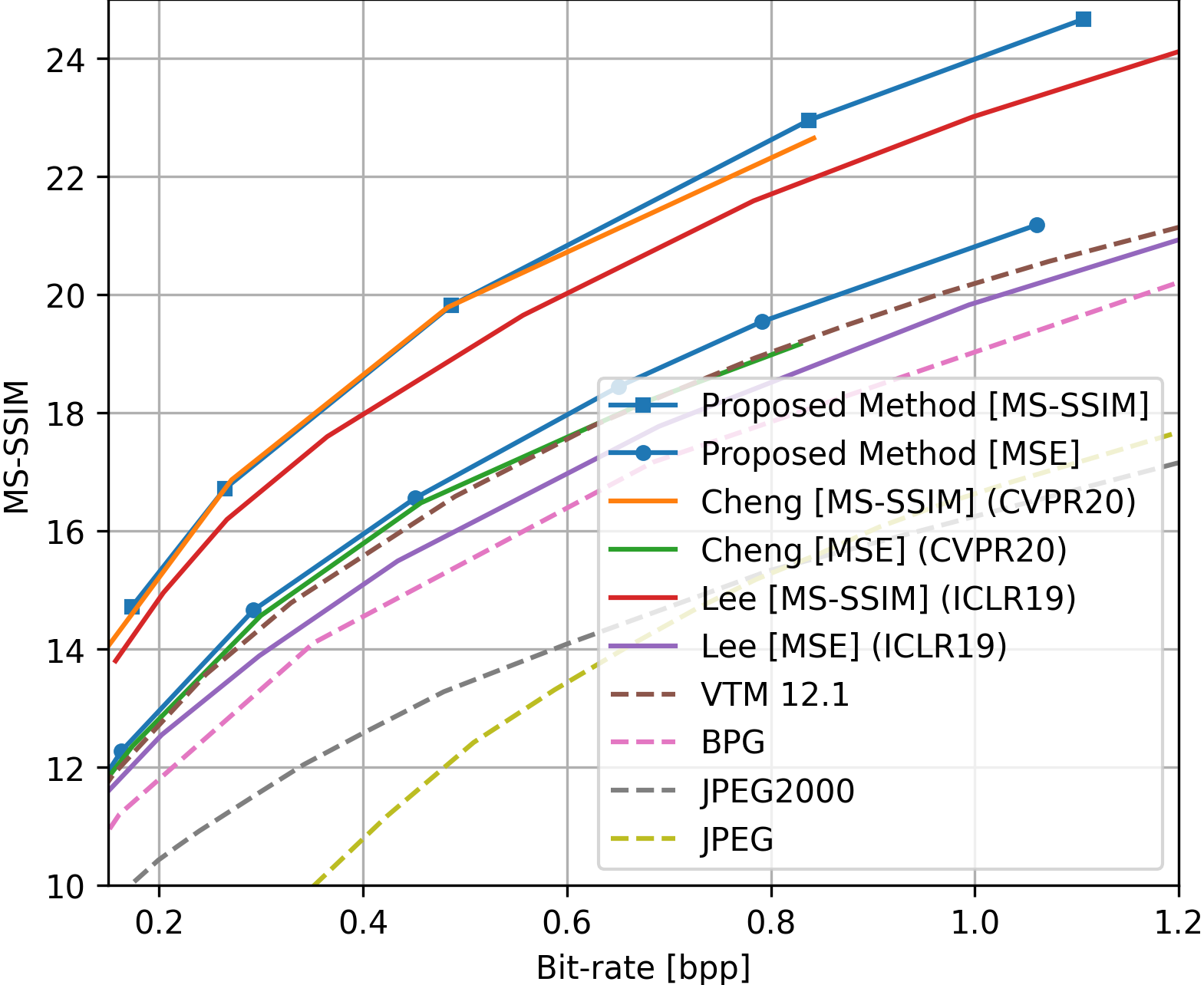}
    \end{subfigure}%
\caption{Performance evaluation on the Kodak dataset.}
\label{fig:result Kodak}
\end{figure*}

\begin{figure}[t]
    \includegraphics[width=1\linewidth]{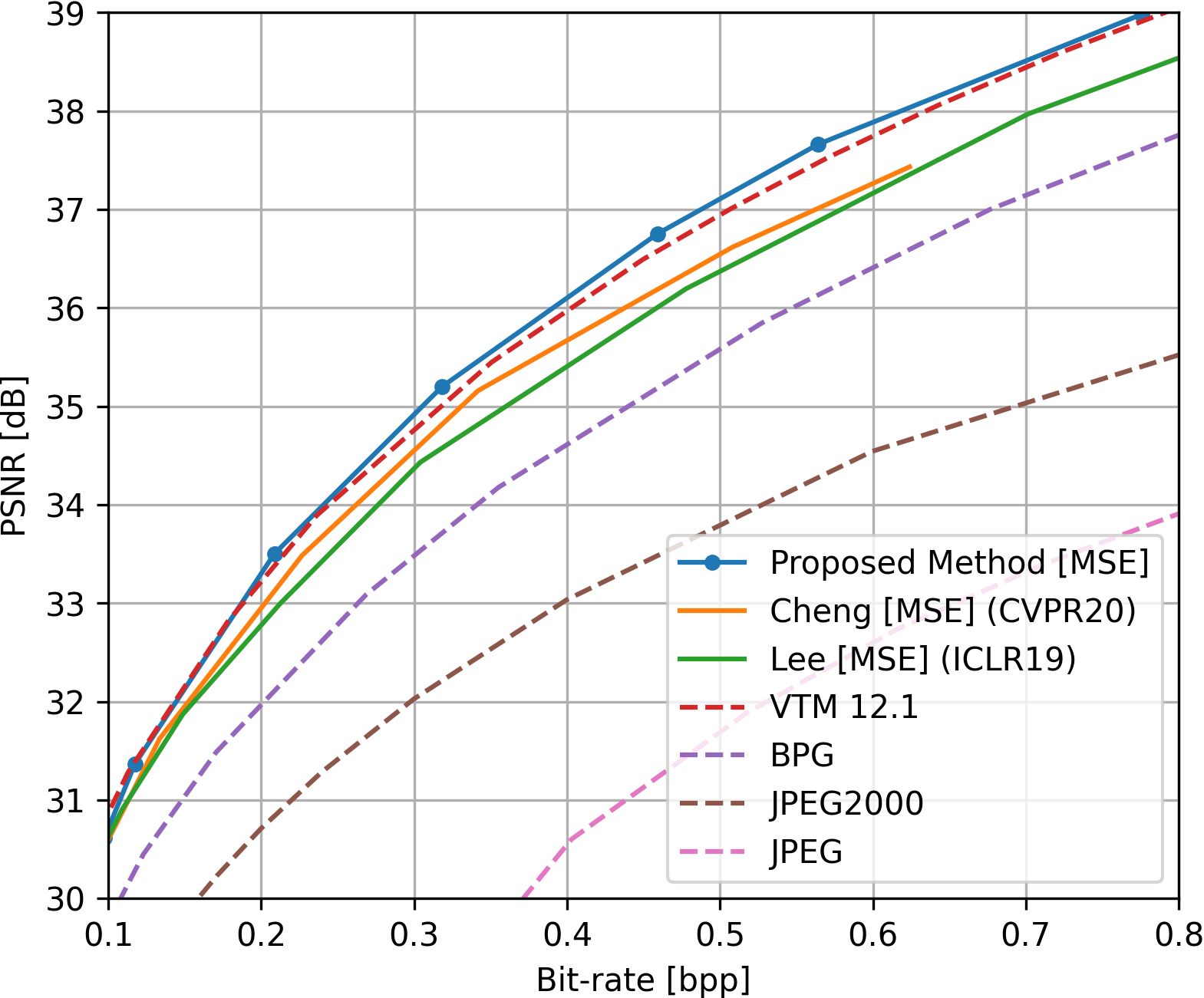}
    \caption{Performance evaluation on the CLIC Professional Validation dataset.}
    \label{fig:result CLIC}
\end{figure}

\begin{figure}[t]
    \includegraphics[width=1\linewidth]{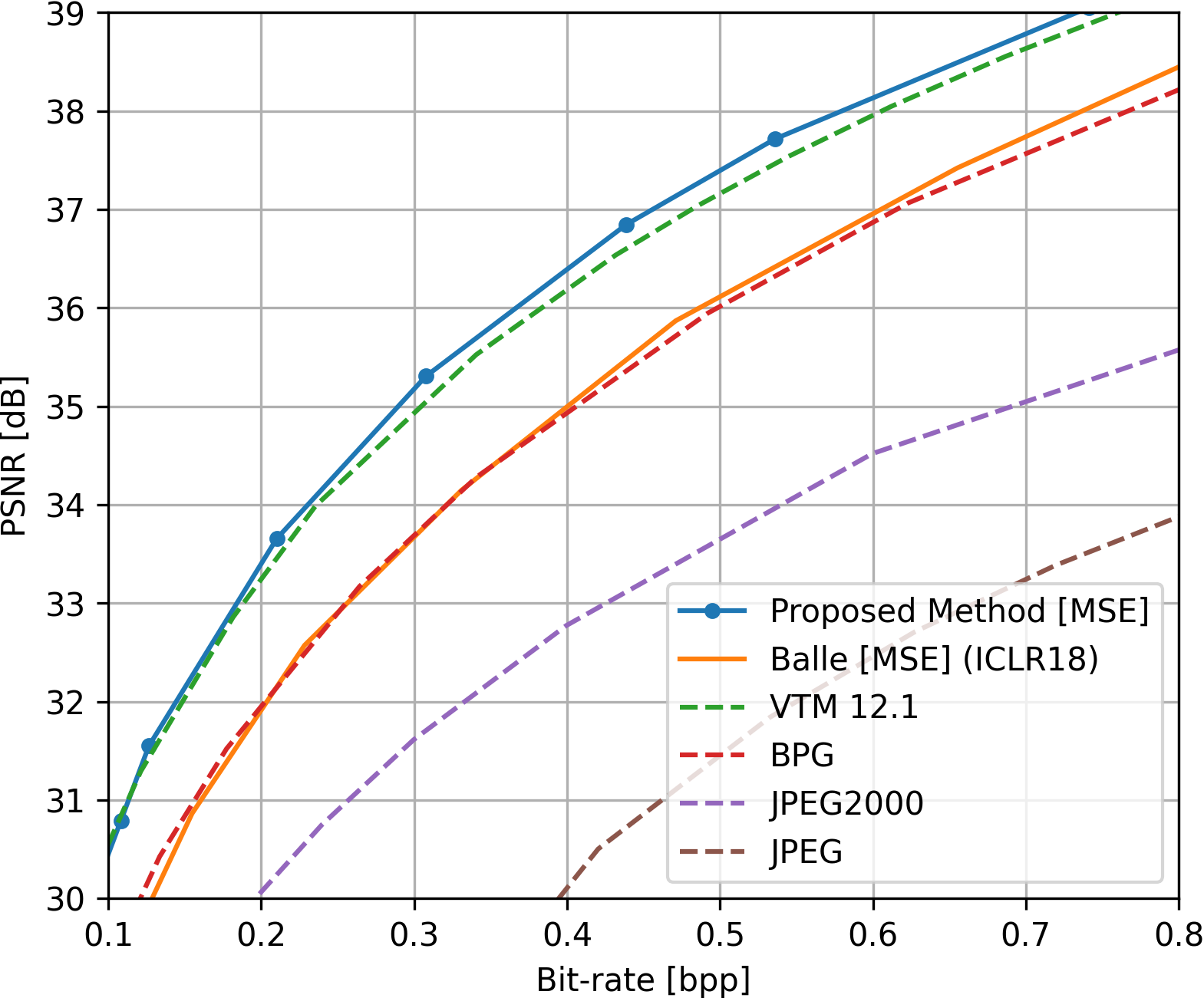}
    \caption{Performance evaluation on the Tecnick dataset.}
    \label{fig:result Tecnick}
\end{figure}

\begin{figure*}[t]
    \begin{subfigure}{1\linewidth}
        \includegraphics[width=1\linewidth]{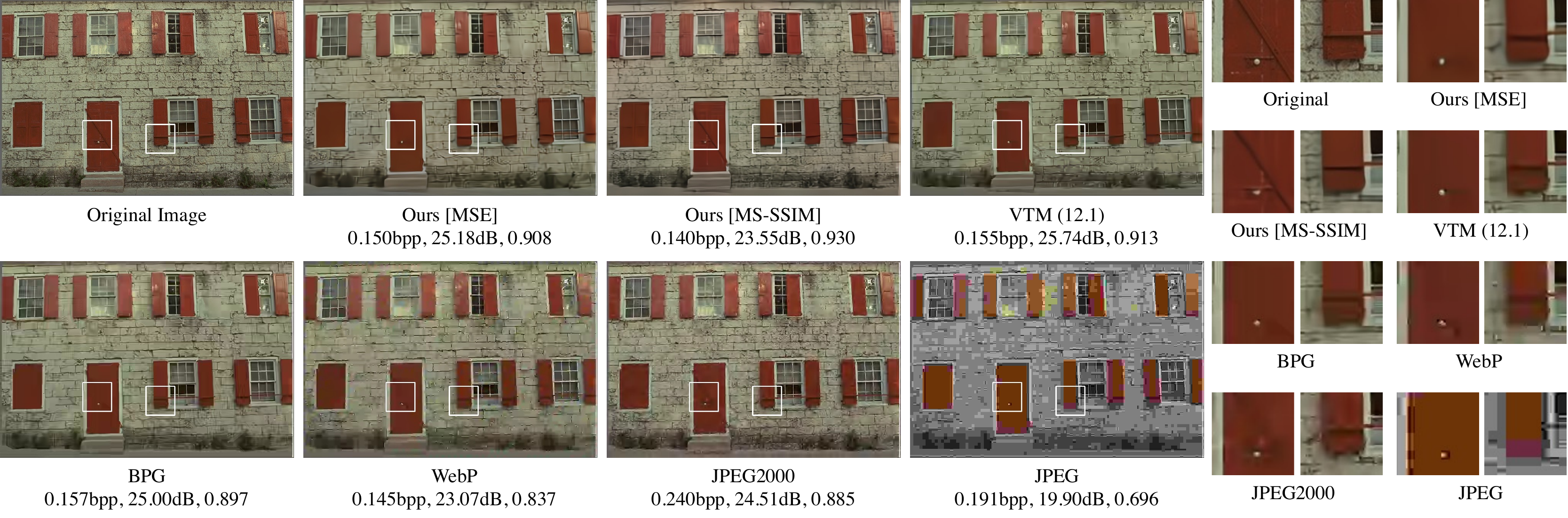}
    \end{subfigure}%
    
    \begin{subfigure}{1\linewidth}
        \includegraphics[width=1\linewidth]{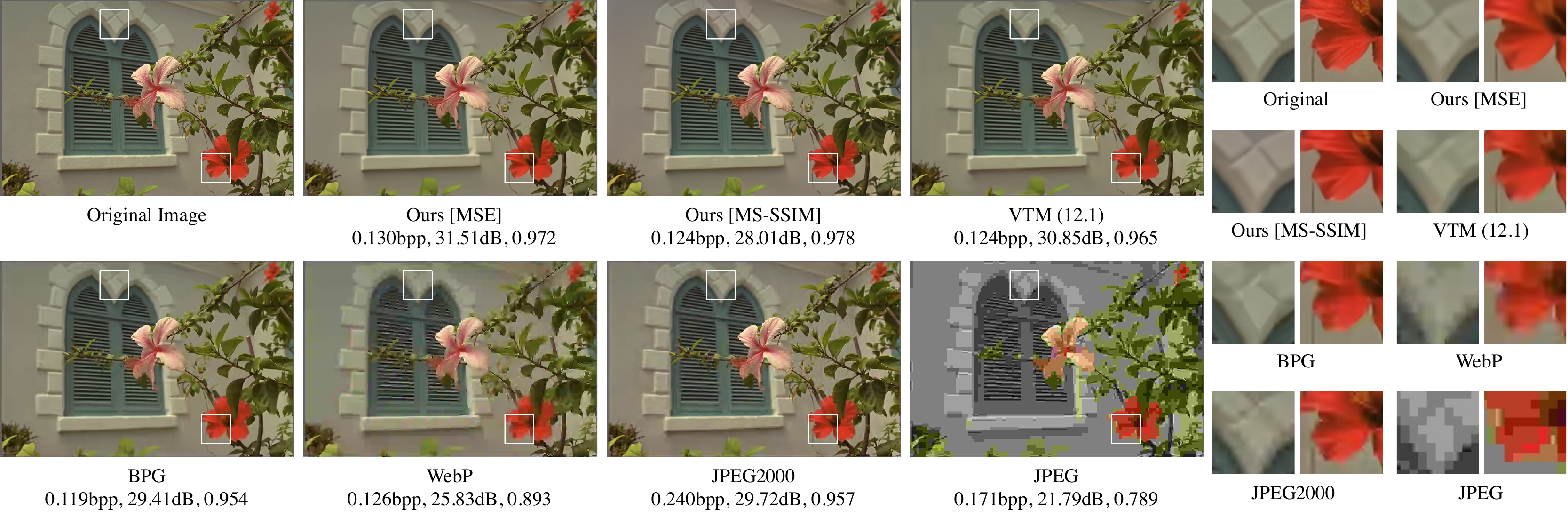}
    \end{subfigure}%
    
    \begin{subfigure}{1\linewidth}
        \includegraphics[width=1\linewidth]{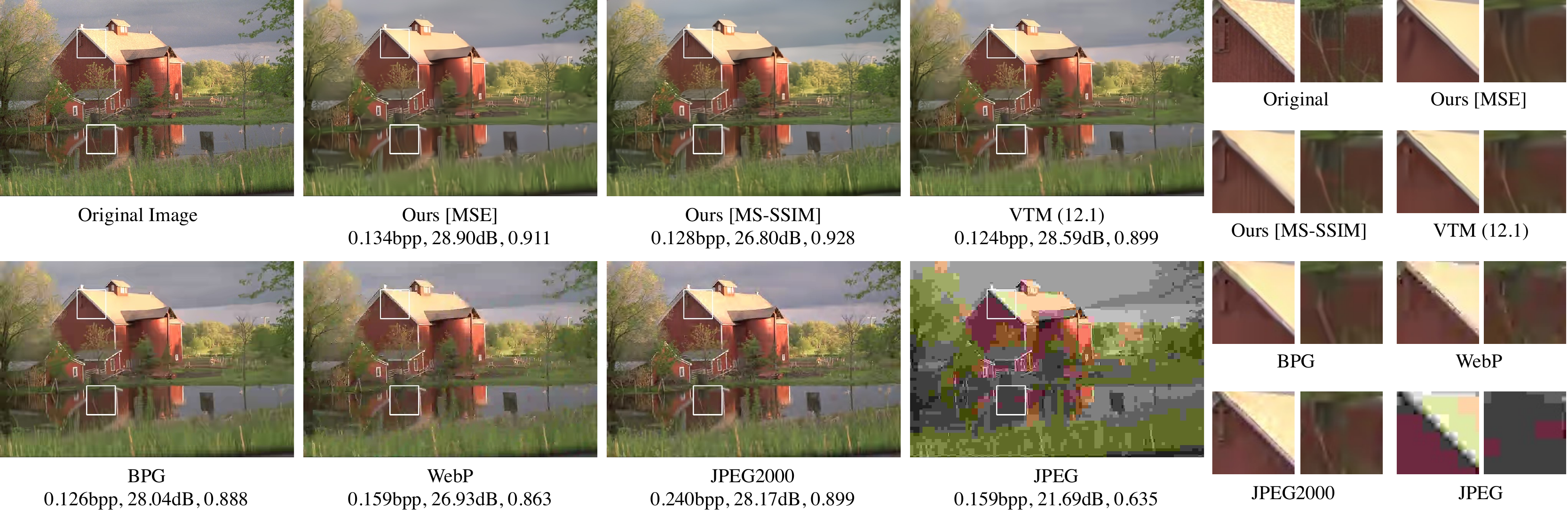}
    \end{subfigure}%
    \caption{Visualization of sample reconstructed images from the Kodak dataset.}
    \label{fig:comparison}
\end{figure*}

\section{Experiments}
We conduct experiments on three commonly used image compression datasets to validate our method. In the remainder of this section, we first introduce the detailed experimental setup, followed by quantitative and qualitative comparisons with existing state-of-the-art methods. We further conduct some analysis and ablation studies to examine the effectiveness of our proposed feature enhancement module and attentive channel squeeze layer.

\subsection{Experimental Setup}
\textbf{Training details.}
We use the Flicker 2W dataset used in~\cite{liu2020unified}, consisting of $20,745$ high-quality general images for the image compression task. We randomly select around $200$ images for our validation set, and the remaining images are used for training. Our network is trained on $256 \times 256$ randomly cropped patches, using the recently well-developed CompressAI PyTorch library~\cite{begaint2020compressai}. Note that we drop a few images with either height or width smaller than $256$ px for convenience.

\begin{table}[t]
  \caption{Area under curve (AUC) of VVC (VTM 12.1) and our MSE method on different datasets, with the bpp range determined by our quality 1 and quality 8 models.}
  \begin{center}
    \begin{tabular}{p{70pt}p{70pt}p{70pt}}
      \toprule[1pt]
      \hfil \textbf{Dataset} & \hfil \textbf{VVC (VTM 12.1)} & \hfil \textbf{Ours [MSE]} \\
      \hline
      \hfil Kodak & \hfil 33.309 & \hfil \textbf{33.373} \\
      \hfil CLIC & \hfil 25.153 & \hfil \textbf{25.238} \\
      \hfil Tecnick & \hfil 23.563 & \hfil \textbf{23.693} \\
      \bottomrule[1pt]
    \end{tabular}
  \end{center}
  \label{tab:mse-auc}
\end{table}

All the experiments are conducted on a single RTX 2080 Ti GPU and trained for $600$ epochs with a batch size of $8$ using Adam~\cite{Kingma2015Adam} optimizer. Generally, it takes around $10$ days to train a model. Our network is first optimized for $450$ epochs with an initial learning rate of $10^{-4}$; then the learning rate is reduced to $10^{-5}$ at epoch $450$ and further down to $10^{-6}$ at epoch $550$.

We use the channel number $N = \frac{C}{\alpha}$ and the weight factor $\lambda$ as our quality parameters. We in-total train $8$ models optimized with MSE (mean squared error) quality metric and $5$ models with MS-SSIM (multiscale structural similarity) quality metric~\cite{wang2003msssim} under different quality levels. For the MSE models, $\lambda$ is chosen from the set $\{0.0016, 0.0024, 0.0032, 0.0075, 0.015, 0.03, 0.045, 0.09\}$, in which the first four $\lambda$ values are paired with channel number $N=128$ for lower-rate models, and $N$ is set as $192$ to pair with the remaining four $\lambda$ values for higher-rate models. For the MS-SSIM models, $\lambda$ belongs to the set of $\{6, 12, 40, 120, 220\}$, and we use $N=128$ for the first two $\lambda$ values to train lower-rate models and $N=192$ for the remaining three $\lambda$ values to train higher-rate models.

\begin{table*}
  \caption{Deviation and scaled deviation of the averaging operation of our attentive channel squeeze layer on the Kodak dataset.}
  \begin{center}
    \begin{tabular}{p{75pt}|p{38pt}p{38pt}p{38pt}p{38pt}p{38pt}p{38pt}p{38pt}p{38pt}}
      \toprule[1pt]
      \hfil \textbf{Quality} & \hfil \textbf{Q1} & \hfil \textbf{Q2} & \hfil \textbf{Q3} & \hfil \textbf{Q4} & \hfil \textbf{Q5} & \hfil \textbf{Q6} & \hfil \textbf{Q7} & \hfil \textbf{Q8} \\
      \hline
      \hfil Deviation ($\epsilon$) & \hfil 0.1344 & \hfil 0.1448 & \hfil 0.1662 & \hfil 0.2344 & \hfil 0.2047 & \hfil 0.2520 & \hfil 0.3541 & \hfil 0.4568 \\
      \hfil Scaled Deviation ($\tilde{\epsilon}$) & \hfil 0.6640 & \hfil 0.5884 & \hfil 0.5912 & \hfil 0.5015 & \hfil 0.4421 & \hfil 0.3637 & \hfil 0.4106 & \hfil 0.3611 \\
      \bottomrule[1pt]
    \end{tabular}
  \end{center}
  \label{tab:analysis-error}
\end{table*}

\begin{figure*}[t]
\begin{subfigure}{1.0\linewidth}
\centering
\includegraphics[width=1.0\linewidth]{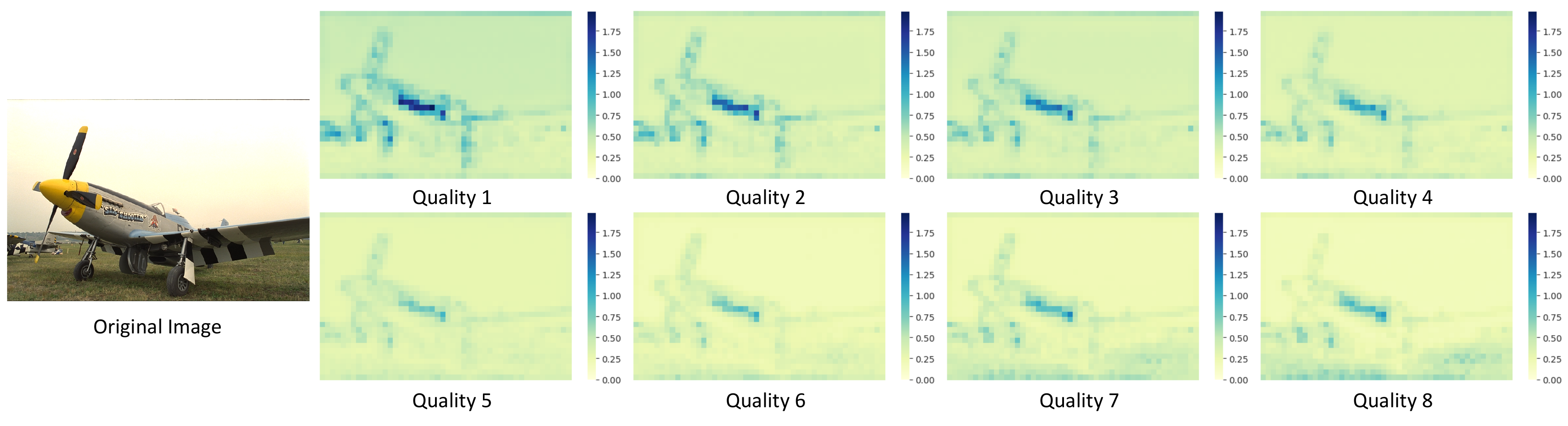}
\end{subfigure}

\begin{subfigure}{1.0\linewidth}
\centering
\includegraphics[width=1.0\linewidth]{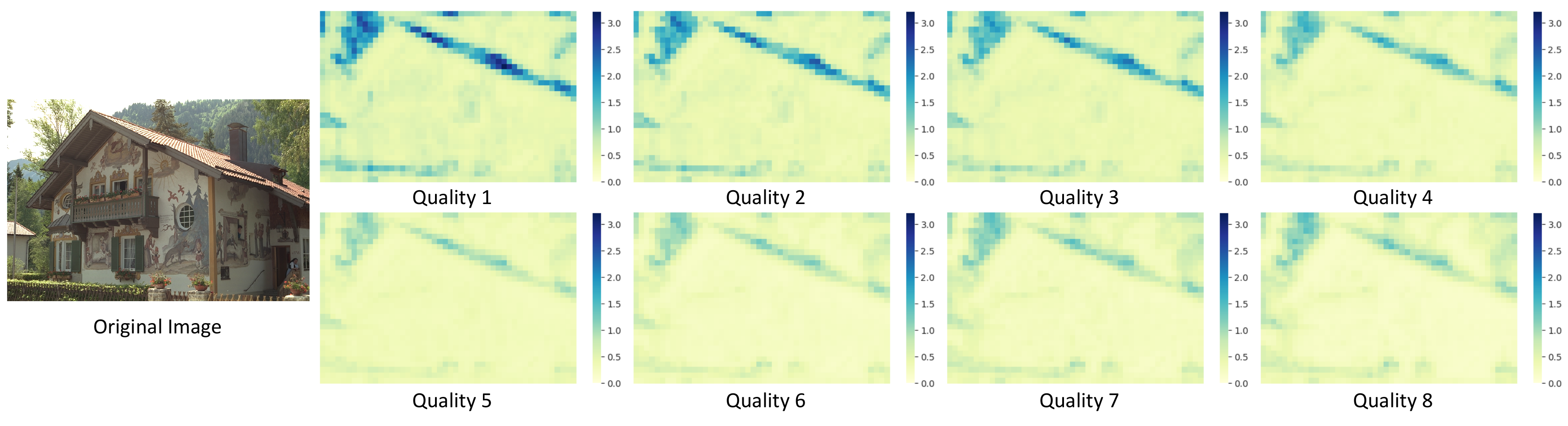}
\end{subfigure}
\caption{Scaled deviation map of image $\textit{kodim20}$ and $\textit{kodim24}$ from the Kodak dataset.}
\label{fig:analysis-error-map}
\end{figure*}

\textbf{Evaluation.} We evaluate our methods on three commonly used datasets for image compression, which are the Kodak PhotoCD image dataset (Kodak) ~\cite{kodak}, the CLIC Professional Validation dataset (CLIC)~\cite{CLIC2020}, and the old Tecnick dataset~\cite{asuni2014tecnick}. The Kodak dataset contains $24$ uncompressed images with resolutions of $768 \times 512$; the CLIC dataset comprises $41$ high-quality images with much higher resolutions; the Tecnick dataset contains $100$ images with high resolutions of $1200 \times 1200$.

We use the peak signal-to-noise ratio (PSNR) and the multiscale structural similarity index (MS-SSIM)~\cite{wang2003msssim} to quantify the image distortion level; we use the bits per pixel (bpp) to evaluate the rate performance. We draw the rate-distortion (RD) curves according to their rate-distortion performance to compare the coding efficiency of different methods. We also report the area under the rate-distortion curve (AUC) as an aggregate measurement to better compare methods with similar performance.

\subsection{Rate-distortion Performance}
We compare our model with state-of-the-art learned image compression models, including the methods proposed by Ball\'{e} el al.~\cite{balle2018variational}, Minnen et al.~\cite{minnen2018joint}, Lee et al.~\cite{lee2018context}, Hu et al.~\cite{hu2020coarse}, and Cheng et al.~\cite{cheng2020learned}. The corresponding data points on their RD curves are collected from their paper or their official GitHub pages. We also compare with some widely-used image compression codecs, including JPEG~\cite{wallace1992jpeg}, JPEG2000~\cite{rabbani2002jpeg2000}, WebP~\cite{webp}, BPG~\cite{bpg}, and VVC~\cite{vvc}. We evaluate their performance using the CompressAI evaluation platform. For VVC, we use the most up-to-date VVC Official Test Model VTM 12.1 (accessed on April 2021)  with an intra-profile configuration from the official GitHub page to test on images. To evaluate the BPG performance, we use the BPG software with subsampling mode of YUV444, HEVC implementation of x265, and bit depth of $8$ to test on images. 

Figure~\ref{fig:result Kodak} shows the RD curve comparison on the Kodak dataset. Similar to existing work~\cite{cheng2020learned}, we convert $MS-SSIM$ to $−10\log_{10}(1−MS\text{-}SSIM)$ for clearer comparison. It can be observed that our method slightly outperforms VVC (VTM 12.1) and yields much better performance when comparing with both the existing learned methods and other traditional image compression standards. 

For the CLIC dataset and the Tecnick dataset, we compare our MSE optimized results with traditional compression standards and the learned methods with official testing results available in their paper or their official GitHub pages. We show the RD curves on the CLIC dataset in Figure~\ref{fig:result CLIC} and the Tecnick dataset in Figure~\ref{fig:result Tecnick}. We can see that our MSE optimized method outperforms all other approaches. Note that most images in the CLIC dataset and the Tecknick dataset are of high resolutions, implying that our method is more robust and promising to compress high-resolution images.

From the RD curves, we can see that VVC (VTM 12.1) and our approach achieve similar performance in terms of PSNR. Hence, we further report their corresponding AUC values for better comparison and ranking. Statistics in Table~\ref{tab:mse-auc} indicates that our method outperforms the latest VVC (VTM 12.1) codec in terms of the aggregated AUC metric.

\subsection{Qualitative Results}
We show some qualitative comparison of some sample reconstructed images on the Kodak dataset in Figure~\ref{fig:comparison}. The first sample is image $\textit{kodim01}$ with an approximate bpp of $0.145$; the second sample is image $\textit{kodim07}$ with approximately $0.125$ bpp; the last sample is image $\textit{kodim22}$ with around $0.130$ bpp. For JPEG and JPEG2000, we use the lowest quality since they cannot reach the mentioned bpp levels. We can see that our MSE optimized method achieves a good performance compared with the latest VVC (VTM 12.1) codec, much better than the performance of other codecs. Also, our MS-SSIM optimized method can reconstruct images with much more structural details than all the traditional codecs. We present more qualitative results in our supplementary materials.

\subsection{Analysis of Attentive Channel Squeeze}
In this section, we demonstrate that our proposed attentive channel squeeze layer introduces only minor deviation to enable a stable and tractable dimension adjustment. We also analyze the distribution of the deviation maps on images and the deviation levels of different quality models.

Let $\mathbf{\gamma} \in \mathbb{R}^{d \times h \times w}$ and $\mathbf{\hat{\gamma}} \in \mathbb{R}^{\frac{d}{\alpha} \times h \times w}$ be the input and output latent feature tensors of the averaging operation, respectively. For simplicity of notation, we reshape the $\mathbf{\gamma}$ as $(\alpha, l)$ and $\mathbf{\hat{\gamma}}$ as $(1, l)$, where $l = \frac{d}{\alpha} \times h \times w$. The averaging operation in the attentive channel squeeze layer compresses $\mathbf{\gamma}$ by compression ratio $\alpha$ along the channel dimension. Specifically, each pixel in $\mathbf{\hat{\gamma}}$ is the mean value of the corresponding $\alpha$ pixels in $\mathbf{\gamma}$. The corresponding inverse operation for dimension matching is copying. It is obvious that such operation can lead to some deviation. 

To quantify such deviation, we do some analysis using the Kodak dataset to calculate the mean absolute pixel deviation as follows:
\begin{equation}
    \epsilon = \frac{1}{l}\sum^{l}_{j=1}{\sum^{\alpha}_{i=1}{|{\gamma_{i,j} - \hat{\gamma}_{1,j}}|}},
\end{equation}
where $\gamma_{i,j}$ and  $\hat{\gamma}_{1,j}$ denote the corresponding pixel in $\mathbf{\gamma}$ and $\mathbf{\hat{\gamma}}$, respectively. As shown in Table~\ref{tab:analysis-error}, the value of the deviation is minor, which is less than or comparable with the error due to the quantization in most cases.

To compare the deviation among models under different quality levels, it is unfair to directly compare the values because $\mathbf{\gamma}$ and $\mathbf{\hat{\gamma}}$ have a larger value range for the model with higher quality. Accordingly, the absolute value of deviation is amplified. As a result, calculating the ``relative'' deviation concerning the value range is a more reasonable practice. A naive way is to divide the mean absolute pixel deviation $\epsilon$ by the value range for fair comparison among different quality models. However, we find that the distribution of the pixel values has long tails, so the value range is very likely to be affected by the outlier pixel values. Hence, to better scale the mean absolute pixel deviation, we introduce a scaling factor $\mu$:
\begin{equation}
    \mu = \frac{1}{l}\sum^{l}_{j=1}{\sum^{\alpha}_{i=1}{|{\gamma_{i,j}|}}}.
\end{equation}
The scaled mean absolute pixel deviation $\tilde{\epsilon} = \frac{\epsilon}{\mu}$. We report the $\epsilon$ and $\tilde{\epsilon}$ values of our models under different quality levels in Table~\ref{tab:analysis-error}.

We further visualize the scaled deviation map of the image $\textit{kodim20}$ and image $\textit{kodim24}$ between $\mathbf{\gamma}$ and $\mathbf{\hat{\gamma}}$ of our models under different quality levels in Figure~\ref{fig:analysis-error-map}. Note that the deviation map is in shape $(h, w)$, and each pixel is the mean of the absolute deviation along the channel dimension after scaling with $\mu$. We can see that the ``relative'' deviation generally decreases as the quality increases, indicating less information loss of the averaging operation, which is in line with our intuition: higher-quality models usually lead to less information loss in the process of compression.

\subsection{Ablation Study} 
To verify the contribution of the proposed nonlinear feature enhancement module, we conduct a corresponding ablation study on this module. Specifically, we train two models with and without using the nonlinear feature enhancement module on the Flicker 2W dataset for $600$ epochs, using the same weight factor $\lambda = 0.01$ and channel number $N = 192$.

\begin{figure}[t]
    \begin{subfigure}{0.5\linewidth}
        \includegraphics[width=1\linewidth]{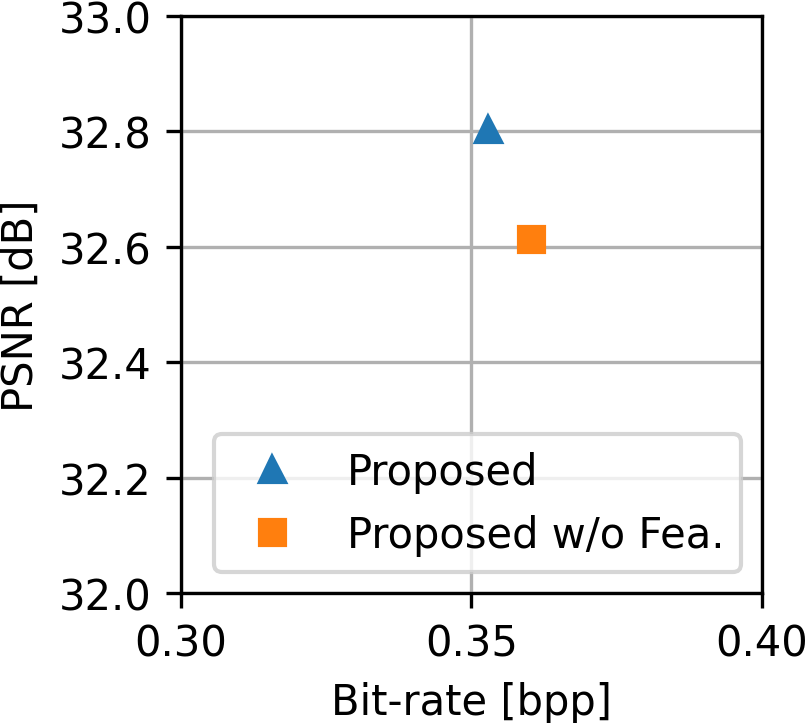}
    \end{subfigure}%
    \begin{subfigure}{0.5\linewidth}
        \includegraphics[width=1\linewidth]{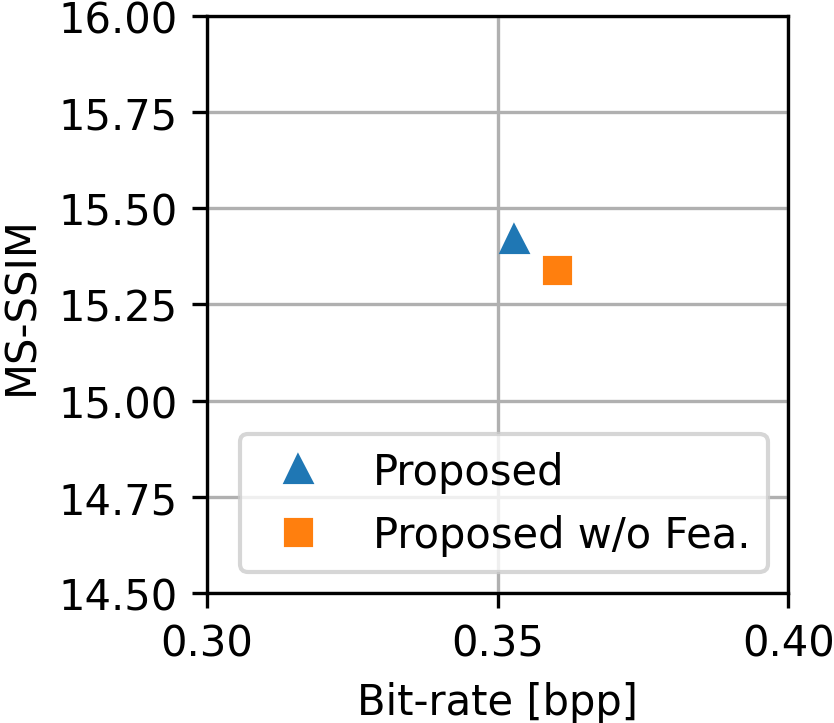}
    \end{subfigure}%
\caption{Ablation study on feature enhancement module.}
\label{fig:ablation1}
\end{figure}

Figure~\ref{fig:ablation1} shows the rate-distortion points of two models evaluated on the Kodak dataset. We can observe that the proposed nonlinear feature enhancement module improves the modeling capacity of the model to compress images with lower bit rates and higher PSNR and MS-SSIM values.

\section{Conclusion}
Unlike existing autoencoder style networks, our proposed enhanced Invertible Encoding Network based on invertible neural networks (INNs) can better model image compression as an invertible process. The critical issues in integrating INNs for image compression include unstable training and limited nonlinear transformation capacity. Our proposed attentive channel squeeze layer offers stable and tractable feature dimension adjustment; the incorporated feature enhancement module increases the network nonlinear transformation capacity. Overall, our network maintains a highly invertible architecture to largely mitigate information loss when compressing images.

Extensive experiments on three widely-used datasets show that our approach outperforms state-of-the-art learned image compression methods and existing compression standards, including VVC (VTM 12.1), especially on two high-resolution image datasets. Furthermore, the visual results demonstrate that our compression methods preserve more detailed information than current compression standards.

\bibliographystyle{ACM-Reference-Format}
\bibliography{compression}


\begin{thebibliography}{52}


\ifx \showCODEN    \undefined \def \showCODEN     #1{\unskip}     \fi
\ifx \showDOI      \undefined \def \showDOI       #1{#1}\fi
\ifx \showISBNx    \undefined \def \showISBNx     #1{\unskip}     \fi
\ifx \showISBNxiii \undefined \def \showISBNxiii  #1{\unskip}     \fi
\ifx \showISSN     \undefined \def \showISSN      #1{\unskip}     \fi
\ifx \showLCCN     \undefined \def \showLCCN      #1{\unskip}     \fi
\ifx \shownote     \undefined \def \shownote      #1{#1}          \fi
\ifx \showarticletitle \undefined \def \showarticletitle #1{#1}   \fi
\ifx \showURL      \undefined \def \showURL       {\relax}        \fi
\providecommand\bibfield[2]{#2}
\providecommand\bibinfo[2]{#2}
\providecommand\natexlab[1]{#1}
\providecommand\showeprint[2][]{arXiv:#2}

\bibitem[\protect\citeauthoryear{Agustsson, Mentzer, Tschannen, Cavigelli,
  Timofte, Benini, and Gool}{Agustsson et~al\mbox{.}}{2017}]%
        {Agustsson2017SofttoHardVQ}
\bibfield{author}{\bibinfo{person}{Eirikur Agustsson}, \bibinfo{person}{Fabian
  Mentzer}, \bibinfo{person}{Michael Tschannen}, \bibinfo{person}{Lukas
  Cavigelli}, \bibinfo{person}{Radu Timofte}, \bibinfo{person}{Luca Benini},
  {and} \bibinfo{person}{Luc~Van Gool}.} \bibinfo{year}{2017}\natexlab{}.
\newblock \showarticletitle{Soft-to-Hard Vector Quantization for End-to-End
  Learning Compressible Representations}. In \bibinfo{booktitle}{\emph{Advances
  in Neural Information Processing Systems}}.
\newblock


\bibitem[\protect\citeauthoryear{Agustsson, Tschannen, Mentzer, Timofte, and
  Gool}{Agustsson et~al\mbox{.}}{2019}]%
        {agustsson2019generative}
\bibfield{author}{\bibinfo{person}{Eirikur Agustsson}, \bibinfo{person}{Michael
  Tschannen}, \bibinfo{person}{Fabian Mentzer}, \bibinfo{person}{Radu Timofte},
  {and} \bibinfo{person}{Luc~Van Gool}.} \bibinfo{year}{2019}\natexlab{}.
\newblock \showarticletitle{Generative Adversarial Networks for Extreme Learned
  Image Compression}. In \bibinfo{booktitle}{\emph{Proceedings of the IEEE/CVF
  International Conference on Computer Vision}}. \bibinfo{pages}{221--231}.
\newblock


\bibitem[\protect\citeauthoryear{Ahmed, Natarajan, and Rao}{Ahmed
  et~al\mbox{.}}{1974}]%
        {ahmed1974dct}
\bibfield{author}{\bibinfo{person}{Nasir~U. Ahmed}, \bibinfo{person}{T.
  Natarajan}, {and} \bibinfo{person}{Kamisetty~R. Rao}.}
  \bibinfo{year}{1974}\natexlab{}.
\newblock \showarticletitle{Discrete Cosine Transform}.
\newblock \bibinfo{journal}{\emph{IEEE Trans. Comput.}} \bibinfo{volume}{C-23},
  \bibinfo{number}{1} (\bibinfo{year}{1974}), \bibinfo{pages}{90--93}.
\newblock


\bibitem[\protect\citeauthoryear{Ardizzone, Kruse, Wirkert, Rahner, Pellegrini,
  Klessen, Maier-Hein, Rother, and K{\"o}the}{Ardizzone et~al\mbox{.}}{2019}]%
        {ardizzone2018analyzing}
\bibfield{author}{\bibinfo{person}{Lynton Ardizzone}, \bibinfo{person}{Jakob
  Kruse}, \bibinfo{person}{Sebastian Wirkert}, \bibinfo{person}{Daniel Rahner},
  \bibinfo{person}{Eric~W. Pellegrini}, \bibinfo{person}{Ralf~S. Klessen},
  \bibinfo{person}{Lena Maier-Hein}, \bibinfo{person}{Carsten Rother}, {and}
  \bibinfo{person}{Ullrich K{\"o}the}.} \bibinfo{year}{2019}\natexlab{}.
\newblock \showarticletitle{Analyzing Inverse Problems with Invertible Neural
  Networks}. In \bibinfo{booktitle}{\emph{Proceedings of the International
  Conference on Learning Representations}}.
\newblock


\bibitem[\protect\citeauthoryear{Asuni and Giachetti}{Asuni and
  Giachetti}{2014}]%
        {asuni2014tecnick}
\bibfield{author}{\bibinfo{person}{Nicola Asuni} {and} \bibinfo{person}{Andrea
  Giachetti}.} \bibinfo{year}{2014}\natexlab{}.
\newblock \showarticletitle{{TESTIMAGES:} a Large-scale Archive for Testing
  Visual Devices and Basic Image Processing Algorithms}. In
  \bibinfo{booktitle}{\emph{Proceedings of the Conference on Smart Tools and
  Applications in Computer Graphics}}. \bibinfo{pages}{63--70}.
\newblock


\bibitem[\protect\citeauthoryear{Ball{\'e}, Laparra, and Simoncelli}{Ball{\'e}
  et~al\mbox{.}}{2016}]%
        {balle2016end}
\bibfield{author}{\bibinfo{person}{Johannes Ball{\'e}}, \bibinfo{person}{Valero
  Laparra}, {and} \bibinfo{person}{Eero~P. Simoncelli}.}
  \bibinfo{year}{2016}\natexlab{}.
\newblock \showarticletitle{End-to-end Optimization of Nonlinear Transform
  Codes for Perceptual Quality}. In \bibinfo{booktitle}{\emph{Proceedings of
  Picture Coding Symposium}}. \bibinfo{pages}{1--5}.
\newblock


\bibitem[\protect\citeauthoryear{Ball{\'e}, Laparra, and Simoncelli}{Ball{\'e}
  et~al\mbox{.}}{2017}]%
        {Ball2017EndtoendOI}
\bibfield{author}{\bibinfo{person}{Johannes Ball{\'e}}, \bibinfo{person}{Valero
  Laparra}, {and} \bibinfo{person}{Eero~P. Simoncelli}.}
  \bibinfo{year}{2017}\natexlab{}.
\newblock \showarticletitle{End-to-end Optimized Image Compression}. In
  \bibinfo{booktitle}{\emph{Proceedings of the International Conference on
  Learning Representations}}.
\newblock


\bibitem[\protect\citeauthoryear{Ball{\'e}, Minnen, Singh, Hwang, and
  Johnston}{Ball{\'e} et~al\mbox{.}}{2018}]%
        {balle2018variational}
\bibfield{author}{\bibinfo{person}{Johannes Ball{\'e}}, \bibinfo{person}{David
  Minnen}, \bibinfo{person}{Saurabh Singh}, \bibinfo{person}{Sung~Jin Hwang},
  {and} \bibinfo{person}{Nick Johnston}.} \bibinfo{year}{2018}\natexlab{}.
\newblock \showarticletitle{Variational Image Compression with a Scale
  Hyperprior}. In \bibinfo{booktitle}{\emph{Proceedings of the International
  Conference on Learning Representations}}.
\newblock


\bibitem[\protect\citeauthoryear{B{\'e}gaint, Racap{\'e}, Feltman, and
  Pushparaja}{B{\'e}gaint et~al\mbox{.}}{2020}]%
        {begaint2020compressai}
\bibfield{author}{\bibinfo{person}{Jean B{\'e}gaint}, \bibinfo{person}{Fabien
  Racap{\'e}}, \bibinfo{person}{Simon Feltman}, {and} \bibinfo{person}{Akshay
  Pushparaja}.} \bibinfo{year}{2020}\natexlab{}.
\newblock \showarticletitle{CompressAI: a PyTorch Library and Evaluation
  Platform for End-to-end Compression Research}.
\newblock \bibinfo{journal}{\emph{arXiv preprint arXiv:2011.03029}}
  (\bibinfo{year}{2020}).
\newblock


\bibitem[\protect\citeauthoryear{Bellard}{Bellard}{2015}]%
        {bpg}
\bibfield{author}{\bibinfo{person}{Fabrice Bellard}.}
  \bibinfo{year}{2015}\natexlab{}.
\newblock \bibinfo{title}{BPG Image Format}.
\newblock
\newblock
\urldef\tempurl%
\url{https://bellard.org/bpg/}
\showURL{%
\tempurl}


\bibitem[\protect\citeauthoryear{Cheng, Sun, Takeuchi, and Katto}{Cheng
  et~al\mbox{.}}{2020}]%
        {cheng2020learned}
\bibfield{author}{\bibinfo{person}{Zhengxue Cheng}, \bibinfo{person}{Heming
  Sun}, \bibinfo{person}{Masaru Takeuchi}, {and} \bibinfo{person}{Jiro Katto}.}
  \bibinfo{year}{2020}\natexlab{}.
\newblock \showarticletitle{Learned Image Compression with Discretized Gaussian
  Mixture Likelihoods and Attention Modules}. In
  \bibinfo{booktitle}{\emph{Proceedings of the IEEE/CVF Conference on Computer
  Vision and Pattern Recognition}}. \bibinfo{pages}{7939--7948}.
\newblock


\bibitem[\protect\citeauthoryear{Company}{Company}{1999}]%
        {kodak}
\bibfield{author}{\bibinfo{person}{Eastman~Kodak Company}.}
  \bibinfo{year}{1999}\natexlab{}.
\newblock \bibinfo{title}{Kodak Lossless True Color Image Suite}.
\newblock
\newblock
\urldef\tempurl%
\url{http://r0k.us/graphics/kodak/}
\showURL{%
\tempurl}


\bibitem[\protect\citeauthoryear{Dinh, Krueger, and Bengio}{Dinh
  et~al\mbox{.}}{2015}]%
        {dinh2015nice}
\bibfield{author}{\bibinfo{person}{Laurent Dinh}, \bibinfo{person}{David
  Krueger}, {and} \bibinfo{person}{Yoshua Bengio}.}
  \bibinfo{year}{2015}\natexlab{}.
\newblock \showarticletitle{NICE: Non-linear Independent Components
  Estimation}. In \bibinfo{booktitle}{\emph{Proceedings of the International
  Conference on Learning Representations Workshops}}.
\newblock


\bibitem[\protect\citeauthoryear{Dinh, Sohl-Dickstein, and Bengio}{Dinh
  et~al\mbox{.}}{2017}]%
        {dinh2017density}
\bibfield{author}{\bibinfo{person}{Laurent Dinh}, \bibinfo{person}{Jascha
  Sohl-Dickstein}, {and} \bibinfo{person}{Samy Bengio}.}
  \bibinfo{year}{2017}\natexlab{}.
\newblock \showarticletitle{Density Estimation using Real NVP}. In
  \bibinfo{booktitle}{\emph{Proceedings of the International Conference on
  Learning Representations}}.
\newblock


\bibitem[\protect\citeauthoryear{Duda}{Duda}{2009}]%
        {duda2009asymmetric}
\bibfield{author}{\bibinfo{person}{Jarek Duda}.}
  \bibinfo{year}{2009}\natexlab{}.
\newblock \showarticletitle{Asymmetric numeral systems}.
\newblock \bibinfo{journal}{\emph{arXiv preprint arXiv:0902.0271}}
  (\bibinfo{year}{2009}).
\newblock


\bibitem[\protect\citeauthoryear{Gersho and Gray}{Gersho and Gray}{2012}]%
        {gersho2012vector}
\bibfield{author}{\bibinfo{person}{Allen Gersho} {and}
  \bibinfo{person}{Robert~M. Gray}.} \bibinfo{year}{2012}\natexlab{}.
\newblock \bibinfo{booktitle}{\emph{Vector Quantization and Signal
  Compression}}. Vol.~\bibinfo{volume}{159}.
\newblock


\bibitem[\protect\citeauthoryear{Google}{Google}{2010}]%
        {webp}
\bibfield{author}{\bibinfo{person}{Google}.} \bibinfo{year}{2010}\natexlab{}.
\newblock \bibinfo{title}{Web Picture Format}.
\newblock
\newblock
\urldef\tempurl%
\url{https://chromium.googlesource.com/webm/libwebp}
\showURL{%
\tempurl}


\bibitem[\protect\citeauthoryear{Goyal}{Goyal}{2001}]%
        {Goyal2001Theoretical}
\bibfield{author}{\bibinfo{person}{Vivek~K. Goyal}.}
  \bibinfo{year}{2001}\natexlab{}.
\newblock \showarticletitle{Theoretical Foundations of Transform Coding}.
\newblock \bibinfo{journal}{\emph{IEEE Signal Processing Magazine}}
  \bibinfo{volume}{18}, \bibinfo{number}{5} (\bibinfo{year}{2001}),
  \bibinfo{pages}{9--21}.
\newblock


\bibitem[\protect\citeauthoryear{Guo, Wu, Feng, Zhang, and Chen}{Guo
  et~al\mbox{.}}{2020}]%
        {guo2020context}
\bibfield{author}{\bibinfo{person}{Zongyu Guo}, \bibinfo{person}{Yaojun Wu},
  \bibinfo{person}{Runsen Feng}, \bibinfo{person}{Zhizheng Zhang}, {and}
  \bibinfo{person}{Zhibo Chen}.} \bibinfo{year}{2020}\natexlab{}.
\newblock \showarticletitle{3-D Context Entropy Model for Improved Practical
  Image Compression}. In \bibinfo{booktitle}{\emph{Proceedings of the IEEE/CVF
  Conference on Computer Vision and Pattern Recognition Workshops}}.
  \bibinfo{pages}{116--117}.
\newblock


\bibitem[\protect\citeauthoryear{Helminger, Djelouah, Gross, and
  Schroers}{Helminger et~al\mbox{.}}{2021}]%
        {helminger2021lossy}
\bibfield{author}{\bibinfo{person}{Leonhard Helminger},
  \bibinfo{person}{Abdelaziz Djelouah}, \bibinfo{person}{Markus Gross}, {and}
  \bibinfo{person}{Christopher Schroers}.} \bibinfo{year}{2021}\natexlab{}.
\newblock \showarticletitle{Lossy Image Compression with Normalizing Flows}. In
  \bibinfo{booktitle}{\emph{Proceedings of the International Conference on
  Learning Representations Workshop Neural Compression}}.
\newblock


\bibitem[\protect\citeauthoryear{Hu, Yang, and Liu}{Hu et~al\mbox{.}}{2020}]%
        {hu2020coarse}
\bibfield{author}{\bibinfo{person}{Yueyu Hu}, \bibinfo{person}{Wenhan Yang},
  {and} \bibinfo{person}{Jiaying Liu}.} \bibinfo{year}{2020}\natexlab{}.
\newblock \showarticletitle{Coarse-to-Fine Hyper-Prior Modeling for Learned
  Image Compression}. In \bibinfo{booktitle}{\emph{Proceedings of the AAAI
  Conference on Artificial Intelligence}}, Vol.~\bibinfo{volume}{34}.
  \bibinfo{pages}{11013--11020}.
\newblock


\bibitem[\protect\citeauthoryear{Huang, Liu, van~der Maaten, and
  Weinberger}{Huang et~al\mbox{.}}{2017}]%
        {huang2017dbnet}
\bibfield{author}{\bibinfo{person}{Gao Huang}, \bibinfo{person}{Zhuang Liu},
  \bibinfo{person}{Laurens van~der Maaten}, {and} \bibinfo{person}{Kilian~Q.
  Weinberger}.} \bibinfo{year}{2017}\natexlab{}.
\newblock \showarticletitle{Densely Connected Convolutional Networks}. In
  \bibinfo{booktitle}{\emph{Proceedings of the IEEE/CVF Conference on Computer
  Vision and Pattern Recognition}}. \bibinfo{pages}{2261--2269}.
\newblock


\bibitem[\protect\citeauthoryear{Johnston, Vincent, Minnen, Covell, Singh,
  Chinen, Hwang, Shor, and Toderici}{Johnston et~al\mbox{.}}{2018}]%
        {johnston2018improved}
\bibfield{author}{\bibinfo{person}{Nick Johnston}, \bibinfo{person}{Damien
  Vincent}, \bibinfo{person}{David Minnen}, \bibinfo{person}{Michele Covell},
  \bibinfo{person}{Saurabh Singh}, \bibinfo{person}{Troy Chinen},
  \bibinfo{person}{Sung~Jin Hwang}, \bibinfo{person}{Joel Shor}, {and}
  \bibinfo{person}{George Toderici}.} \bibinfo{year}{2018}\natexlab{}.
\newblock \showarticletitle{Improved Lossy Image Compression with Priming and
  Spatially Adaptive Bit Rates for Recurrent Networks}. In
  \bibinfo{booktitle}{\emph{Proceedings of the IEEE Conference on Computer
  Vision and Pattern Recognition}}. \bibinfo{pages}{4385--4393}.
\newblock


\bibitem[\protect\citeauthoryear{(JVET)}{(JVET)}{2021}]%
        {vvc}
\bibfield{author}{\bibinfo{person}{Joint Video Experts~Team (JVET)}.}
  \bibinfo{year}{2021}\natexlab{}.
\newblock \bibinfo{title}{VVC Official Test Model VTM}.
\newblock
\newblock
\urldef\tempurl%
\url{https://vcgit.hhi.fraunhofer.de/jvet/VVCSoftware_VTM/-/tree/VTM-12.1}
\showURL{%
\tempurl}
\newblock
\shownote{accessed on April 5, 2021.}


\bibitem[\protect\citeauthoryear{Kingma and Ba}{Kingma and Ba}{2015}]%
        {Kingma2015Adam}
\bibfield{author}{\bibinfo{person}{Diederik~P. Kingma} {and}
  \bibinfo{person}{Jimmy Ba}.} \bibinfo{year}{2015}\natexlab{}.
\newblock \showarticletitle{Adam: {A} Method for Stochastic Optimization}. In
  \bibinfo{booktitle}{\emph{Proceddings of the International Conference on
  Learning Representations}}.
\newblock


\bibitem[\protect\citeauthoryear{Kingma and Dhariwal}{Kingma and
  Dhariwal}{2018}]%
        {kingma2018glow}
\bibfield{author}{\bibinfo{person}{Durk~P. Kingma} {and}
  \bibinfo{person}{Prafulla Dhariwal}.} \bibinfo{year}{2018}\natexlab{}.
\newblock \showarticletitle{Glow: Generative Flow with Invertible 1x1
  Convolutions}. In \bibinfo{booktitle}{\emph{Advances in Neural Information
  Processing Systems}}, Vol.~\bibinfo{volume}{31}.
\newblock


\bibitem[\protect\citeauthoryear{Lee, Cho, and Beack}{Lee
  et~al\mbox{.}}{2019}]%
        {lee2018context}
\bibfield{author}{\bibinfo{person}{Jooyoung Lee}, \bibinfo{person}{Seunghyun
  Cho}, {and} \bibinfo{person}{Seung-Kwon Beack}.}
  \bibinfo{year}{2019}\natexlab{}.
\newblock \showarticletitle{Context-adaptive Entropy Model for End-to-end
  Optimized Image Compression}. In \bibinfo{booktitle}{\emph{Proceedings of the
  International Conference on Learning Representations}}.
\newblock


\bibitem[\protect\citeauthoryear{Lin, Yao, Chen, and Wang}{Lin
  et~al\mbox{.}}{2020}]%
        {lin2020}
\bibfield{author}{\bibinfo{person}{Chaoyi Lin}, \bibinfo{person}{Jiabao Yao},
  \bibinfo{person}{Fangdong Chen}, {and} \bibinfo{person}{Li Wang}.}
  \bibinfo{year}{2020}\natexlab{}.
\newblock \showarticletitle{A Spatial RNN Codec for End-to-End Image
  Compression}. In \bibinfo{booktitle}{\emph{Proceedings of the IEEE/CVF
  Conference on Computer Vision and Pattern Recognition}}.
  \bibinfo{pages}{13266--13274}.
\newblock


\bibitem[\protect\citeauthoryear{Liu, Chen, Guo, Shen, Cao, Wang, and Ma}{Liu
  et~al\mbox{.}}{2019}]%
        {liu2019non}
\bibfield{author}{\bibinfo{person}{Haojie Liu}, \bibinfo{person}{Tong Chen},
  \bibinfo{person}{Peiyao Guo}, \bibinfo{person}{Qiu Shen},
  \bibinfo{person}{Xun Cao}, \bibinfo{person}{Yao Wang}, {and}
  \bibinfo{person}{Zhan Ma}.} \bibinfo{year}{2019}\natexlab{}.
\newblock \showarticletitle{Non-local Attention Optimized Deep Image
  Compression}.
\newblock \bibinfo{journal}{\emph{arXiv preprint arXiv:1904.09757}}
  (\bibinfo{year}{2019}).
\newblock


\bibitem[\protect\citeauthoryear{Liu, Lu, Hu, and Xu}{Liu
  et~al\mbox{.}}{2020}]%
        {liu2020unified}
\bibfield{author}{\bibinfo{person}{Jiaheng Liu}, \bibinfo{person}{Guo Lu},
  \bibinfo{person}{Zhihao Hu}, {and} \bibinfo{person}{Dong Xu}.}
  \bibinfo{year}{2020}\natexlab{}.
\newblock \showarticletitle{A Unified End-to-End Framework for Efficient Deep
  Image Compression}.
\newblock \bibinfo{journal}{\emph{arXiv preprint arXiv:2002.03370}}
  (\bibinfo{year}{2020}).
\newblock


\bibitem[\protect\citeauthoryear{Lugmayr, Danelljan, Gool, and Timofte}{Lugmayr
  et~al\mbox{.}}{2020}]%
        {lugmayr2020srflow}
\bibfield{author}{\bibinfo{person}{Andreas Lugmayr}, \bibinfo{person}{Martin
  Danelljan}, \bibinfo{person}{Luc~Van Gool}, {and} \bibinfo{person}{Radu
  Timofte}.} \bibinfo{year}{2020}\natexlab{}.
\newblock \showarticletitle{SRFlow: Learning the Super-Resolution Space with
  Normalizing Flow}. In \bibinfo{booktitle}{\emph{Proceedings of the European
  Conference on Computer Vision}}.
\newblock


\bibitem[\protect\citeauthoryear{Marpe, Schwarz, and Wiegand}{Marpe
  et~al\mbox{.}}{2003}]%
        {marpe2003context}
\bibfield{author}{\bibinfo{person}{Detlev Marpe}, \bibinfo{person}{Heiko
  Schwarz}, {and} \bibinfo{person}{Thomas Wiegand}.}
  \bibinfo{year}{2003}\natexlab{}.
\newblock \showarticletitle{Context-based Adaptive Binary Arithmetic Coding in
  the H. 264/AVC Video Compression Standard}.
\newblock \bibinfo{journal}{\emph{IEEE Transactions on Circuits and Systems for
  Video Technology}} \bibinfo{volume}{13}, \bibinfo{number}{7}
  (\bibinfo{year}{2003}), \bibinfo{pages}{620--636}.
\newblock


\bibitem[\protect\citeauthoryear{Mentzer, Agustsson, Tschannen, Timofte, and
  Gool}{Mentzer et~al\mbox{.}}{2018}]%
        {mentzer2018conditional}
\bibfield{author}{\bibinfo{person}{Fabian Mentzer}, \bibinfo{person}{Eirikur
  Agustsson}, \bibinfo{person}{Michael Tschannen}, \bibinfo{person}{Radu
  Timofte}, {and} \bibinfo{person}{Luc~Van Gool}.}
  \bibinfo{year}{2018}\natexlab{}.
\newblock \showarticletitle{Conditional Probability Models for Deep Image
  Compression}. In \bibinfo{booktitle}{\emph{Proceedings of the IEEE Conference
  on Computer Vision and Pattern Recognition}}. \bibinfo{pages}{4394--4402}.
\newblock


\bibitem[\protect\citeauthoryear{Minnen, Ball{\'e}, and Toderici}{Minnen
  et~al\mbox{.}}{2018}]%
        {minnen2018joint}
\bibfield{author}{\bibinfo{person}{David Minnen}, \bibinfo{person}{Johannes
  Ball{\'e}}, {and} \bibinfo{person}{George Toderici}.}
  \bibinfo{year}{2018}\natexlab{}.
\newblock \showarticletitle{Joint Autoregressive and Hierarchical Priors for
  Learned Image Compression}. In \bibinfo{booktitle}{\emph{Advances in Neural
  Information Processing Systems}}. \bibinfo{pages}{10794--10803}.
\newblock


\bibitem[\protect\citeauthoryear{Minnen and Singh}{Minnen and Singh}{2020}]%
        {MinnenS20}
\bibfield{author}{\bibinfo{person}{David Minnen} {and} \bibinfo{person}{Saurabh
  Singh}.} \bibinfo{year}{2020}\natexlab{}.
\newblock \showarticletitle{Channel-Wise Autoregressive Entropy Models for
  Learned Image Compression}. In \bibinfo{booktitle}{\emph{Proceedings of the
  IEEE International Conference on Image Processing}}.
  \bibinfo{pages}{3339--3343}.
\newblock


\bibitem[\protect\citeauthoryear{Pumarola, Popov, Moreno-Noguer, and
  Ferrari}{Pumarola et~al\mbox{.}}{2020}]%
        {pumarola2020cflow}
\bibfield{author}{\bibinfo{person}{Albert Pumarola}, \bibinfo{person}{Stefan
  Popov}, \bibinfo{person}{Francesc Moreno-Noguer}, {and}
  \bibinfo{person}{Vittorio Ferrari}.} \bibinfo{year}{2020}\natexlab{}.
\newblock \showarticletitle{C-Flow: Conditional Generative Flow Models for
  Images and 3D Point Clouds}. In \bibinfo{booktitle}{\emph{Proceedings of the
  IEEE/CVF Conference on Computer Vision and Pattern Recognition}}.
\newblock


\bibitem[\protect\citeauthoryear{Rabbani}{Rabbani}{2002}]%
        {rabbani2002jpeg2000}
\bibfield{author}{\bibinfo{person}{Majid Rabbani}.}
  \bibinfo{year}{2002}\natexlab{}.
\newblock \showarticletitle{JPEG2000: Image Compression Fundamentals, Standards
  and Practice}.
\newblock \bibinfo{journal}{\emph{Journal of Electronic Imaging}}
  \bibinfo{volume}{11}, \bibinfo{number}{2} (\bibinfo{year}{2002}),
  \bibinfo{pages}{286}.
\newblock


\bibitem[\protect\citeauthoryear{Rippel and Bourdev}{Rippel and
  Bourdev}{2017}]%
        {rippel2017real}
\bibfield{author}{\bibinfo{person}{Oren Rippel} {and} \bibinfo{person}{Lubomir
  Bourdev}.} \bibinfo{year}{2017}\natexlab{}.
\newblock \showarticletitle{Real-time Adaptive Image Compression}. In
  \bibinfo{booktitle}{\emph{Proceedings of the International Conference on
  Machine Learning}}. \bibinfo{pages}{2922--2930}.
\newblock


\bibitem[\protect\citeauthoryear{Rissanen and Langdon}{Rissanen and
  Langdon}{1981}]%
        {Rissanen1981Universal}
\bibfield{author}{\bibinfo{person}{Jorma Rissanen} {and}
  \bibinfo{person}{Glen~G. Langdon}.} \bibinfo{year}{1981}\natexlab{}.
\newblock \showarticletitle{Universal Modeling and Coding}.
\newblock \bibinfo{journal}{\emph{IEEE Transactions on Information Theory}}
  \bibinfo{volume}{27}, \bibinfo{number}{1} (\bibinfo{year}{1981}),
  \bibinfo{pages}{12--23}.
\newblock


\bibitem[\protect\citeauthoryear{Santurkar, Budden, and Shavit}{Santurkar
  et~al\mbox{.}}{2018}]%
        {santurkar2017generative}
\bibfield{author}{\bibinfo{person}{Shibani Santurkar},
  \bibinfo{person}{David~M. Budden}, {and} \bibinfo{person}{Nir Shavit}.}
  \bibinfo{year}{2018}\natexlab{}.
\newblock \showarticletitle{Generative Compression}. In
  \bibinfo{booktitle}{\emph{Proceedings of Picture Coding Symposium}}.
  \bibinfo{pages}{258--262}.
\newblock


\bibitem[\protect\citeauthoryear{Theis, Shi, Cunningham, and Husz{\'a}r}{Theis
  et~al\mbox{.}}{2017}]%
        {Theis2017LossyIC}
\bibfield{author}{\bibinfo{person}{Lucas Theis}, \bibinfo{person}{Wenzhe Shi},
  \bibinfo{person}{Andrew Cunningham}, {and} \bibinfo{person}{Ferenc
  Husz{\'a}r}.} \bibinfo{year}{2017}\natexlab{}.
\newblock \showarticletitle{Lossy Image Compression with Compressive
  Autoencoders}. In \bibinfo{booktitle}{\emph{Proceedings of the International
  Conference on Learning Representations}}.
\newblock


\bibitem[\protect\citeauthoryear{Toderici, O'Malley, Hwang, Vincent, Minnen,
  Baluja, Covell, and Sukthankar}{Toderici et~al\mbox{.}}{2015}]%
        {toderici2015variable}
\bibfield{author}{\bibinfo{person}{George Toderici}, \bibinfo{person}{Sean~M.
  O'Malley}, \bibinfo{person}{Sung~Jin Hwang}, \bibinfo{person}{Damien
  Vincent}, \bibinfo{person}{David Minnen}, \bibinfo{person}{Shumeet Baluja},
  \bibinfo{person}{Michele Covell}, {and} \bibinfo{person}{Rahul Sukthankar}.}
  \bibinfo{year}{2015}\natexlab{}.
\newblock \showarticletitle{Variable Rate Image Compression with Recurrent
  Neural Networks}. In \bibinfo{booktitle}{\emph{Proceedings of the
  International Conference on Learning Representations}}.
\newblock


\bibitem[\protect\citeauthoryear{Toderici, Shi, Timofte, Lucas~Theis,
  Agustsson, Johnston, and Mentzer}{Toderici et~al\mbox{.}}{2021}]%
        {CLIC2020}
\bibfield{author}{\bibinfo{person}{George Toderici}, \bibinfo{person}{Wenzhe
  Shi}, \bibinfo{person}{Radu Timofte}, \bibinfo{person}{Johannes~Balle
  Lucas~Theis}, \bibinfo{person}{Eirikur Agustsson}, \bibinfo{person}{Nick
  Johnston}, {and} \bibinfo{person}{Fabian Mentzer}.}
  \bibinfo{year}{2021}\natexlab{}.
\newblock \bibinfo{title}{Workshop and Challenge on Learned Image Compression}.
\newblock
\newblock
\urldef\tempurl%
\url{http://www.compression.cc}
\showURL{%
\tempurl}


\bibitem[\protect\citeauthoryear{Toderici, Vincent, Johnston, Hwang, Minnen,
  Shor, and Covell}{Toderici et~al\mbox{.}}{2017}]%
        {toderici2017full}
\bibfield{author}{\bibinfo{person}{George Toderici}, \bibinfo{person}{Damien
  Vincent}, \bibinfo{person}{Nick Johnston}, \bibinfo{person}{Sung~Jin Hwang},
  \bibinfo{person}{David Minnen}, \bibinfo{person}{Joel Shor}, {and}
  \bibinfo{person}{Michele Covell}.} \bibinfo{year}{2017}\natexlab{}.
\newblock \showarticletitle{Full Resolution Image Compression with Recurrent
  Neural Networks}. In \bibinfo{booktitle}{\emph{Proceedings of the IEEE
  Conference on Computer Vision and Pattern Recognition}}.
  \bibinfo{pages}{5306--5314}.
\newblock


\bibitem[\protect\citeauthoryear{Wallace}{Wallace}{1992}]%
        {wallace1992jpeg}
\bibfield{author}{\bibinfo{person}{Gervais~Knox Wallace}.}
  \bibinfo{year}{1992}\natexlab{}.
\newblock \showarticletitle{The JPEG Still Picture Compression Standard}.
\newblock \bibinfo{journal}{\emph{IEEE Transactions on Consumer Electronics}}
  \bibinfo{volume}{38}, \bibinfo{number}{1} (\bibinfo{year}{1992}),
  \bibinfo{pages}{xviii--xxxiv}.
\newblock


\bibitem[\protect\citeauthoryear{Wang, Xiao, Liu, Zheng, and Liu}{Wang
  et~al\mbox{.}}{2020}]%
        {wang2020modeling}
\bibfield{author}{\bibinfo{person}{Yaolong Wang}, \bibinfo{person}{Mingqing
  Xiao}, \bibinfo{person}{Chang Liu}, \bibinfo{person}{Shuxin Zheng}, {and}
  \bibinfo{person}{Tie-Yan Liu}.} \bibinfo{year}{2020}\natexlab{}.
\newblock \showarticletitle{Modeling Lost Information in Lossy Image
  Compression}.
\newblock \bibinfo{journal}{\emph{arXiv preprint arXiv:2006.11999}}
  (\bibinfo{year}{2020}).
\newblock


\bibitem[\protect\citeauthoryear{Wang, Simoncelli1, and Bovik}{Wang
  et~al\mbox{.}}{2003}]%
        {wang2003msssim}
\bibfield{author}{\bibinfo{person}{Zhou Wang}, \bibinfo{person}{Eero~P.
  Simoncelli1}, {and} \bibinfo{person}{Alan~C. Bovik}.}
  \bibinfo{year}{2003}\natexlab{}.
\newblock \showarticletitle{Multiscale Structural Similarity for Image Quality
  Assessment}. In \bibinfo{booktitle}{\emph{Proceedings of the Asilomar
  Conference on Signals, Systems and Computers}}, Vol.~\bibinfo{volume}{2}.
  \bibinfo{pages}{1398--1402}.
\newblock


\bibitem[\protect\citeauthoryear{Witten, Neal, and Cleary}{Witten
  et~al\mbox{.}}{1987}]%
        {arith}
\bibfield{author}{\bibinfo{person}{Ian~H. Witten}, \bibinfo{person}{Radford~M.
  Neal}, {and} \bibinfo{person}{John~G. Cleary}.}
  \bibinfo{year}{1987}\natexlab{}.
\newblock \showarticletitle{Arithmetic Coding for Data Compression}.
\newblock \bibinfo{journal}{\emph{Commun. ACM}} \bibinfo{volume}{30},
  \bibinfo{number}{6} (\bibinfo{year}{1987}), \bibinfo{pages}{520–540}.
\newblock


\bibitem[\protect\citeauthoryear{Xiao, Zheng, Liu, Wang, He, Ke, Bian, Lin, and
  Liu}{Xiao et~al\mbox{.}}{2020}]%
        {xiao2020invertible}
\bibfield{author}{\bibinfo{person}{Mingqing Xiao}, \bibinfo{person}{Shuxin
  Zheng}, \bibinfo{person}{Chang Liu}, \bibinfo{person}{Yaolong Wang},
  \bibinfo{person}{Di He}, \bibinfo{person}{Guolin Ke}, \bibinfo{person}{Jiang
  Bian}, \bibinfo{person}{Zhouchen Lin}, {and} \bibinfo{person}{Tie-Yan Liu}.}
  \bibinfo{year}{2020}\natexlab{}.
\newblock \showarticletitle{Invertible Image Rescaling}. In
  \bibinfo{booktitle}{\emph{Proceedings of the European Conference on Computer
  Vision}}. \bibinfo{pages}{126--144}.
\newblock


\bibitem[\protect\citeauthoryear{Xing, Qian, and Chen}{Xing
  et~al\mbox{.}}{2021}]%
        {Xing2021}
\bibfield{author}{\bibinfo{person}{Yazhou Xing}, \bibinfo{person}{Zian Qian},
  {and} \bibinfo{person}{Qifeng Chen}.} \bibinfo{year}{2021}\natexlab{}.
\newblock \showarticletitle{Invertible Image Signal Processing}. In
  \bibinfo{booktitle}{\emph{Proceedings of the IEEE/CVF International
  Conference on Computer Vision}}.
\newblock


\bibitem[\protect\citeauthoryear{Zhang, Li, Li, Zhong, and Fu}{Zhang
  et~al\mbox{.}}{2019}]%
        {zhang2019residual}
\bibfield{author}{\bibinfo{person}{Yulun Zhang}, \bibinfo{person}{Kunpeng Li},
  \bibinfo{person}{Kai Li}, \bibinfo{person}{Bineng Zhong}, {and}
  \bibinfo{person}{Yun Fu}.} \bibinfo{year}{2019}\natexlab{}.
\newblock \showarticletitle{Residual Non-local Attention Networks for Image
  Restoration}. In \bibinfo{booktitle}{\emph{Proceedings of the International
  Conference on Learning Representations}}.
\newblock


\bibitem[\protect\citeauthoryear{Zhou, Sun, Wu, and Wu}{Zhou
  et~al\mbox{.}}{2019}]%
        {zhou2019end}
\bibfield{author}{\bibinfo{person}{Lei Zhou}, \bibinfo{person}{Zhenhong Sun},
  \bibinfo{person}{Xiangji Wu}, {and} \bibinfo{person}{Junmin Wu}.}
  \bibinfo{year}{2019}\natexlab{}.
\newblock \showarticletitle{End-to-end Optimized Image Compression with
  Attention Mechanism.}. In \bibinfo{booktitle}{\emph{Proceedings of the IEEE
  Conference on Computer Vision and Pattern Recognition Workshops}}.
\newblock


\end{thebibliography}


\end{document}